\documentclass[fleqn,usenatbib]{mnras}

\usepackage{amssymb,bm}
\usepackage{subfigure}
\usepackage{bm} 
\usepackage{graphicx}
\usepackage{float}
\usepackage{amsmath}
\usepackage{url}
\usepackage[usenames,dvipsnames,svgnames,table]{xcolor}
\usepackage{color}
\definecolor{color1}{RGB}{191, 0, 255}

\def \mnras {MNRAS}
\title{Magnetic winding and turbulence in ultra-hot Jupiters}
\author[Soriano-Guerrero \& al.]{Cl\`audia Soriano-Guerrero$^{1,2}$\thanks{E-mail: soriano@ice.csic.es}, Daniele Vigan\`o$^{1,2,3}$, Rosalba Perna$^{4,5}$,
Taner Akg\"un$^{1}$,
\newauthor
Carlos Palenzuela$^{6,2,3}$
\\
$^1$Institute of Space Sciences (ICE, CSIC), 08193 Barcelona, Spain\\
$^2$Institut d'Estudis Espacials de Catalunya (IEEC), 08034 Barcelona, Spain\\
$^3$Institut Aplicacions Computationals (IAC3),  Universitat  de  les  Illes  Balears,  Palma  de  Mallorca,  Baleares  E-07122,  Spain\\
$^4$Department of Physics and Astronomy, Stony Brook University, Stony Brook, NY 11794-3800, USA\\
$^5$Center for Computational Astrophysics, Flatiron Institute, New York, NY 10010, USA\\
$^6$Departament  de  F\'{\i}sica,  Universitat  de  les  Illes  Balears,  Palma  de  Mallorca,  Baleares  E-07122,  Spain
}

\date{}

\pubyear{2023}

\begin{document}
\label{firstpage}
\pagerange{\pageref{firstpage}--\pageref{lastpage}}
\maketitle

\begin{abstract}
While magnetism in exoplanets remains largely unknown, Hot Jupiters have been considered as natural candidates to harbour intense magnetic fields, both due to their large masses which might empower a larger internal dynamo, and, possibly, due to their high energy budgets coming from irradiation. In this work we focus on the latter aspect and perform MHD simulations of a narrow day-side atmospheric column of ultra-hot Jupiters, 
suitable
for very high local temperatures ($T \gtrsim 3000$ K). Due to the high conductivity in this regime, the primary influence is the winding of the magnetic field caused by the intense zonal winds. In our study, we include a forcing that mimics the wind profiles observed in  GCMs near the sub-stellar point. As a result, the shear layer generates a toroidal magnetic field, locally reaching few kG, which is supported by meridional currents. Such fields and the sustaining currents do not depend on the internal field, but are all confined in the thin (less than a scale-height) shear layer around 1 bar. Additionally, we add random perturbations that induce turbulent motions, which lead to further (but much smaller) magnetic field generation to a broader range of depths. These results enable the assessment of the atmospheric currents that are induced.
Although here we use ideal MHD and the only resistivity comes from the numerical scheme at a fixed resolution, we estimate a-posteriori the amount of Ohmic heat deposited in the outer layers, which could be employed in evolutionary models for Hot Jupiters' inflated radii.

\end{abstract}


\section{Introduction}\label{sec:intro}

Hot Jupiters (HJs, see e.g. \cite{heng15,fortney21} for reviews) are a peculiar class of exoplanets: gas giants orbiting very close to their host stars, with orbital periods of a few days, corresponding to distances within $\sim 0.1$~AU. Due to the high irradiation from their host stars and the tidal locking of the rotational and orbital periods, strong temperature differences persist between the dayside and the nightside, as proved also by direct observations (e.g. \citealt{knutson07}). Such temperature gradients generate strong zonal jets that try to redistribute the heat (e.g., \citealt{cho08,dobbs08,showman09,heng11,rauscher12,perna12,rauscher13,perez13,parmentier13,rogers14b,showman15,kataria15,koll18,beltz22,komacek22,dietrich22}).

A feature of these hot giants is that most of them have inflated radii, reaching up to $R \sim 2 R_j$ (Jovian radii), which cannot be accounted for within standard cooling models for planetary evolution (e.g. \citealt{showman02,wang15}). A strong correlation with the strength of the irradiation (or equivalently the equilibrium temperature of the planet, $T_{eq}$) is observed, with inflation seen to begin above about $T_{eq}\sim 1000$~K (e.g., \citealt{laughlin11}, or Fig.~1 of \citealt{thorngren18}).\footnote{The equilibrium temperature is defined as $T_{eq}=T_\star(R_\star/a)^{1/2}(1-A)^{1/4}$, where $A$ is the planetary albedo, $a$ is the orbital star-planet separation, and $R_\star$ and $T_\star$ are the star's radius and temperature, respectively.} This indicates a continuous deposition of heat, corresponding to up to a few percent of the irradiation flux\citep{komacek17}, that slows down their long-term cooling, or even re-inflate them \citep{lopez16,komacek20}. For these reasons, other mechanisms that would slow down internal cooling regardless of the irradiation (increased opacity, \citealt{burrows07}, double-diffusive convection, \citealt{chabrier07}) are considered less favoured to be the dominant ones.

Although the correlation is evident, the effects of irradiation alone are not quantitatively sufficient to explain the large radii inflation, given the shallowness of the absorbing layer of the radiation.
Various possible heating mechanisms have been proposed \citep{bodenheimer01,chabrier07,li10,youdin10,batygin10,perna10b}. Among them, a popular one is Ohmic dissipation \citep{batygin10,perna10b}, produced by the dissipation of currents induced by the magnetic field stretching due to the flow motion \citep{liu08,perna10a}. Such currents are a natural outcome of the strong height-dependent winds composed of ionized material. Statistical studies of HJ populations \citep{thorngren18} show how the inferred relation between heat mechanism efficiency shows a peak at $T_{eq}\sim 1500-1600$ K. This naturally fits the Ohmic scenario, since two competing effects can lower the heating efficiency as the temperature increases: on one side, the wind and, therefore, the induced magnetic fields are stronger; on the other side, magnetic drag slows down more the thermal winds \citep{perna10a,rogers14a} and the resistivity decreases due to the increasing thermal ionization \citep{kumar21}.

In order to quantify the amount of such atmospheric Ohmic dissipation, we indeed need a prescription for both the conductivity and currents radial profiles.
In this sense, 3D magnetohydrodynamic (MHD) simulations \citep{batygin13,rogers14a,rogers14b}, or estimations of the magnetic drag and Ohmic timescales \citep{rauscher13,beltz22}, have been performed within global circulation models (GCMs, see \citealt{heng15} for a review). In possibly the most advanced work in this respect so far, \cite{rogers17} demonstrated that a dynamo can be maintained in the thin, stably-stratified atmosphere of the planet, driven by spatial conductivity variations resulting from the asymmetric temperature profile in the planet.

On the other hand, such global simulations cannot capture small-scale turbulence, due to lack of resolution. Turbulence in the HJ atmospheres has been indeed investigated separately in a number of works, but with little attention on the magnetic field. \cite{li10} analyzed the possibility of generating turbulence in circulation models. They performed a linear analysis and 2D hydrodynamics simulations under the presence of laminar forcing in the horizontal direction. They found that turbulent and vertical mixing are dominant dissipation mechanisms, as a result of a recurrent Kelvin-Helmholtz instability triggered by the zonal winds. \cite{youdin10} assumed the presence of turbulent flows and studied semi-analytically the effects of the turbulent diffusion and energy deposition on the radiative equilibrium profile of the atmosphere. They suggested a so-called mechanical greenhouse effect, with the turbulence driving heat at inner atmospheric depths. \cite{ryu18} used box simulations to quantify the possibility of generating turbulence via the shear layer caused by the strong zonal winds in the upper atmosphere. Their simulations were however in the purely hydrodynamic regime, and hence no magnetic field induction could be evaluated.

Complementary, other studies \citep{liu08,batygin10,perna10b,kumar21} considered a semi-analytical estimation of the induced magnetic field. They operated in the framework of the linear regime of the induction, i.e. when the ionized wind produces a negligible perturbation over the background magnetic field generated in the interior. This approach is valid for relatively low conductivities, corresponding to temperatures $T\lesssim 1500$ K \citep{dietrich22} at most. For higher temperatures, the induced magnetic field is locally high enough that the linear approximation fails.

Here, we face the problem from another yet complementary angle: a quantitative assessment of the MHD induction in the non-linear regime, i.e. when the induced magnetic fields are comparable or larger than the internal ones. For the first time to the best of our knowledge, we present local MHD simulations of a narrow atmospheric column in the dayside radiative layers of a HJ upper atmosphere (range of pressure $\sim$ mbar-10 bar). Our aim is to have a first assessment of the combined effects of winding and turbulence  under some simplified assumptions on the background state and resistivity. On one side, we simulate the generation of strong toroidal fields in the wind shear layer, as also pointed out very recently by \cite{dietrich22}, after we started this project. 
We include realistic, parametrized profiles for the wind velocity, qualitatively mimicking the steepest profiles of GCMs \citep{beltz22}. On the other side, we enforce turbulent perturbations in addition to the zonal wind. We aim at evaluating where electrical currents are induced, and quantify how the wind profile affects the generation of currents and local magnetic fields. This is a first step, since the resistivity is given here implicitly by the numerical scheme at a fixed resolution, rather than by the physical resistivity.

The paper is structured as follows.
In \S\ref{sec:equations} we formulate the general problem and the equations we solve. In \S\ref{sec:setup} we give details on the numerical infrastructure and methods. In \S\ref{sec:results} we present the results and discuss their applicability, and we provide a rough estimate of the local Ohmic dissipation in \S\ref{sec:applicability}. We finally draw the main conclusions in \S\ref{sec:conclusions}.

\section{Model of a hot Atmospheric Column}\label{sec:equations}

\subsection{Equation of state and hydrostatic equilibrium}

We assume an ideal gas equation of state (EoS):
\begin{equation}\label{eq:eos}
p = \rho R T = (\gamma - 1)\rho e~,
\end{equation}
where $p$ is the gas pressure, $\rho$ the mass density, $e$ the specific internal energy (energy per unit mass), $T$ the temperature, $\gamma$ the ideal gas index (fixed to 1.4 in this work), $R=k_b/m = (8.254/\mu)\times 10^3$ J/(kg\,K) the specific gas constant, with $k_b$ the Boltzmann constant, $m=\mu m_u$ the mean molecular mass of the gas and $m_u$ the atomic mass unit. The mean molecular weight $\mu$ depends on the chemical composition and on the degree of ionization. HJ atmospheres are arguably dominated by Hydrogen, Helium and traces of heavier atoms and molecules; as a reference, in the Solar System, atmospheres of ice/gas giants have $\mu \sim (2-2.7)$. However, such values are lower in HJs due to the dissociation of molecules, $H_2$ in particular (see e.g. Fig. 1 of \citealt{parmentier18}), so we can consider $\mu$ of order one.

Under the presence of gravity directed downwards in the $z$ direction, $\vec{g}=-g \hat{z}$, the equation of hydrostatic equilibrium reads
\begin{equation}\label{eq:hydro_eq}
\frac{dp}{dz} = -\rho g ~.
\end{equation}
We can safely assume $g$ to be constant in the domain of interest, since the atmospheric mass and thickness are much smaller than the enclosed mass and the planetary radius, respectively. As shown in analytical and numerical studies (e.g., \citealt{guillot10,youdin10}), the outermost HJ radiative layers (well above the radiative-conductive boundary, RCB) are approximately isothermal. Here we consider a localized, narrow column of the central dayside with a uniform background temperature $T_0$.

Then the background profile is simply:
\begin{equation}\label{eq:p_exp}
   p(z) = p_b e^{-z/H_*}~,
\end{equation}
where $p_b$ is the pressure at $z=0$, and we have introduced the pressure scale-height, defined as $H_* \equiv ||p/\vec{\nabla}p||$, which in this case means: \begin{equation}\label{eq:def_hp}
H_* \equiv \frac{R T_0}{g} = 1.66\times 10^6 {\rm m} ~\frac{T_{2000}}{\mu~g_{10}}~,
\end{equation}
where $T_{2000}=T/(2000$ K) and $g_{10}=g/(10$ m~s$^{-2}$).

The value of $T_0$ (which is assigned a-posteriori due to the scale invariance of the equations, as discussed below) is higher than $T_{eq}$, if we consider a column close to the substellar point, because the temperature varies with longitude (e.g., \citealt{rogers14b}), unless the thermal redistribution is unrealistically perfect.

\subsection{Ideal compressible MHD equations}

Under the typical conditions of a HJ upper atmosphere, $T\sim 1000$-$3000$ K, its gas contains partially ionized species, in particular K$^+$ and other alkaline metals \citep{parmentier18,kumar21,dietrich22}. Therefore, the presence of strong zonal winds implies the circulation of electrical currents. If they are strong enough, they can produce a magnetic field which can be locally comparable or higher than the one  generated in the interior.

Our aim is to find the steady (but not static) state of a magnetized, turbulent and strongly stratified atmosphere, in the presence of a zonal wind shear. This is achieved by adding a forcing term (as a proxy for a slightly turbulent zonal wind) and leaving the quantities to evolve until they statistically stabilize. At such a stationary state, the two dominant terms in the induction equation roughly balance each other: advection and Ohmic dissipation. Additional Hall and ambipolar terms are estimated to be negligible, at least for magnetic fields of the order of a few Gauss \citep{perna10a}. 

Our simulations will be testing 
the non-linear regime (as defined by \citealt{dietrich22}), i.e. with high conductivity and magnetic Reynolds number, ${\cal R}m = \mu_0 \sigma v L \gg 1$, where $\sigma$ is the electrical conductivity, $\mu_0=1.2566\times10^{-6}$ kg m/(s$^2$A$^2$) is the vacuum magnetic permeability, and $v$ and $L$ are the typical velocity and length-scale of the fluid. In the non-linear regime, the winding effect is expected to create a strong toroidal field, which is dynamically relevant and more intense than the background field. In other words, the physical value of resistivity is expected to be pretty low, and the field can wind up considerably, supported by localized currents.

In this first work we use ideal MHD equations, neglecting the explicit inclusion of thermal, viscous or magnetic diffusivities. In reality, dissipative effects are implicitly given by our numerical scheme. For very high temperatures ($T\gtrsim 3000$ K) and the resolution used here, the numerical and physical magnetic diffusivities are comparable (see below and Appendix~\ref{app:diffusivity} for details). Therefore, the results presented here should not be far from a non-ideal MHD case, which will be the subject of a follow-up work. However, the magnetic fields could be slightly lower than in this study.

Using the Einstein notation, where the sub/superscripts $i,j,k$ indicate the three spatial components and the repeated indexes indicate a sum over components, and a fully conservative formulation, indicating with $[~\cdot~]$ the fluxes and leaving the sources at the right-hand side, the equations read:

\begin{eqnarray}
\partial_t\rho &+& \partial_k \left[ \rho v^k \right] = 0 ~,\label{eq:mhd} \\
\partial_t S^i &+& \partial_k \left[ \rho v^k v^i - \frac{B^k B^i}{\mu_0} + \delta^{ki}\left(p + \frac{B^2}{2\mu_0}\right) \right] = \nonumber \\
&& = \rho g^i + F^i_{\rm src} ~, \\
\partial_t B^i &+& \partial_k \left[ v^k B^i - v^i B^k  \right] = 0 ~, \\
\partial_t E &+& \partial_k \left[\left(E  + p + \frac{B^2}{2\mu_0}\right) v^k -(v_j B^j) \frac{B^k}{\mu_0} \right] = \nonumber\\
&& =  \rho v_j g^j + v_j F^j_{\rm src} ~, \label{eq:mhd4}
\end{eqnarray}
where $\delta^{ki}$ is the Kronecker delta, and we have introduced the momentum $S^i=\rho v^i$, the fluid velocity $v^i$, the magnetic field $B^i$, and the total energy density $E$ (total energy per unit volume), defined as
\begin{equation}\label{eq:energy_definition}
    E = \rho e + \frac{1}{2}\left(\rho v^2 + \frac{B^2}{\mu_0}\right) = \frac{p}{\gamma - 1} + \frac{1}{2}\left(\rho v^2 + \frac{B^2}{\mu_0}\right)~.
\end{equation}
We choose the Cartesian system of reference $(x,y,z)$ so that the gravity has components $g^i = (0,0,-g)$. The conserved quantities $\{ \rho, S^i, B^i, E\}$ are numerically evolved via equations discretized in space and time. At each step of a simulation, the velocity is recovered simply by $v^i=S^i/\rho$ and the pressure by inverting eq.~(\ref{eq:energy_definition}):
\begin{equation}\label{eq:p_recovery}
    p = (\gamma - 1)\left[E - \frac{1}{2}\left(\rho v^2 + \frac{B^2}{\mu_0}\right)\right]~.
\end{equation}
The other thermodynamical quantities (e.g., $e,T$) can be obtained via eq. (\ref{eq:eos}). 
The source term $F^i_{\rm src}$ represents any external force, which appears in the momentum and energy equation (see \S~\ref{sec:forcing} for the specific prescription of the forcing term).
The recovered variables $v^i,p$ (or $e$, or $T$, instead of $p$), together with $\rho$ and $B^i$, are usually called ``primitive'' in MHD: they constitute the complete, basic set of fields and thermodynamical quantities that describe the system.

\subsection{Rescaled equations}\label{sec:rescaled_eqs}

The MHD equations can be rescaled so that the coordinates and all the fields become dimensionless as follows (variables are denoted with $\hat{\cdot}$, reference values with $\cdot_*$; here, $x$ is representative of any spatial coordinate):
\begin{eqnarray}
    && \rho = \rho_* \hat{\rho},~~
    p = p_* \hat{p},~~
    v = c_* \hat{v},~~ B = B_* \hat{B} ~, \\
    && x = H_* \hat{x} ~\Rightarrow~
    \partial_x
    = \frac{1}{H_*} \partial_{\hat x}~,\\
    && t = t_* \hat{t} ~\Rightarrow~
    \partial_t
    = \frac{1}{t_*} \partial_{\hat t}~.
\end{eqnarray}
When one of the reference values is fixed (e.g., $p_*$), the remaining ones are conveniently written as:
\begin{eqnarray}
    && B_* := \sqrt{\mu_0 p_*} \simeq 0.354 \sqrt{p_{\rm bar}}~{\rm T}~,\\
    && \rho_* := \frac{p_*}{RT_0} \simeq 6.01\cdot 10^{-3} \frac{\rm kg}{{\rm m}^3}~ \frac{p_{\rm bar} \mu}{T_{2000}}~,\\
    && c_* := \sqrt{\frac{p_*}{\rho_*}} = \sqrt{R T_0} \simeq 4.08\times 10^3\frac{\rm m}{\rm s} \sqrt{\frac{T_{2000}}{\mu}}~, \\
    && t_* := \frac{H_*}{c_*} = \frac{R T_0}{g c_*} = \frac{c_*}{g} \simeq 408~{\rm s} \frac{\sqrt{T_{2000}}}{g_{10}\sqrt{{\mu}}}~,
\end{eqnarray}
where $p_{\rm bar} := p_*/(1$ bar) and $H_*(g,T_0)$ is given by eq.~ (\ref{eq:def_hp}). Note that $c_{s,0}=\sqrt{\gamma p/\rho}=\sqrt{\gamma}c_*$ is the speed of sound at the given $T_0$ and $\mu$, and $t_{x,0}=H_*/c_{s,0}=t_*/\sqrt{\gamma}$ is the related crossing timescale for one scale-height. Note also that $1/t_* = (\sqrt{ \gamma}/2) \nu_{c,0}$, where $\nu_{c,0}=\gamma g/(2 c_{s,0})$ is the cut-off frequency of the acoustic waves in a stratified atmosphere (e.g. \citealt{tobias_thesis}, eq. 2.66).

With this convenient definition of the six reference values $(H_*, p_*, \rho_*, c_*, B_*, t_*)$, the MHD eqs. (\ref{eq:mhd})-(\ref{eq:mhd4}) look the same, replacing the original variables with the dimensionless ones (i.e., the values used in the code), and reabsorbing $\mu_0$:
\begin{eqnarray}
\partial_{\hat{t}}\hat{\rho} &+& \partial_{\hat{k}} \left[ \hat{\rho} \hat{v}^k \right] = 0 ~, \label{eq:mhdnorm} \\
\partial_{\hat{t}}\hat{S}^i &+& \partial_{\hat{k}} \left[ \hat{\rho} \hat{v}^k \hat{v}^i - \hat{B}^k \hat{B}^i + \delta^{ki}\left(\hat{p} + \frac{\hat{B}^2}{2}\right) \right] = \nonumber \\
&& = \hat{\rho} \hat{g}^i  + \hat{F}^i_{\rm src}  ~, \label{eq:mhdnorm2} \\
\partial_{\hat{t}} \hat{B}^i &+& \partial_{\hat{k}} \left[ \hat{v}^k \hat{B}^i - \hat{v}^i \hat{B}^k  \right] = 0 ~, \label{eq:mhdnorm3} \\
\partial_{\hat{t}} \hat{E} &+& \partial_{\hat{k}} \left[\left(\hat{E}  + \hat{p} + \frac{\hat{B}^2}{2}\right) \hat{v}^k -(\hat{v}_j \hat{B}^j) \hat{B}^k \right] = \nonumber\\
&& = \hat{\rho} \hat{v}_j \hat{g}^i + \hat{v}_j \hat{F}^j_{\rm src} 
 ~, \label{eq:mhdnorm4} 
\end{eqnarray}
where the rescaled definitions of gravity, momentum, total energy density, specific internal energy, sources and equation of state are simply
\begin{eqnarray}
    \hat{g}^i &=& g^i/g=(0,0,-1)~,\\
    \hat{S}^i &=& \hat{\rho} \hat{v}^i ~, \\
    \hat{E} &=& \frac{E}{p_*} = \hat{\rho}\hat{e} + \frac{1}{2}(\hat{\rho} \hat{v}^2 + \hat{B}^2)~, \\
    \hat{e} &=& \frac{e}{c_*^2} ~, \\
    \hat{F}_{\rm src} &=& \frac{F_{\rm src}}{\rho_* g}~,\label{eq:forcingnorm}\\
    \hat{p} &=& \hat{\rho}\hat{T} = (\gamma - 1)\hat{\rho}\hat{e}~,\label{eq:eos_rescaled}
\end{eqnarray}
where $\hat{T}=T/T_0$ is the dimensionless temperature. Finally, the electrical currents, $\vec{J}=(\vec{\nabla}\times\vec{B})/ \mu_0$, can be similarly redefined as $\vec{J}= J_*\vec{\hat{J}}= J_*(\vec{\hat{\nabla}}\times\vec{\hat{B}})$, where the conversion factor is
\begin{equation}\label{eq:jstar}
    J_* = \frac{B_*}{\mu_0 H_*} \simeq 0.17 {\rm \frac{A}{m^2}} ~ \frac{\mu ~g_{10} \sqrt{p_{\rm bar}}}{T_{2000}}~. 
\end{equation}
In this work, we keep simulations scale-invariant, i.e., we are agnostic over the set of values ($p_*,g,\mu,T_0$), which are assigned a-posteriori in order to convert the dimensionless values (i.e., code units) into physical quantities.

\subsection{Background-perturbation approach}

As common in numerical simulations of  strongly stratified media, we decompose the field in a static background component, $\cdot_0$ (for which the time evolution is always zero), plus a perturbed one, $\cdot_1$, for instance:
\begin{eqnarray}
   p &=& p_0 + p_1~, \label{eq:pdec} \\
   \rho &=& \rho_0 + \rho_1~. \label{eq:rhodec}
\end{eqnarray}
In particular, for the atmosphere the background component is suitably represented by the stable, stratified structure at a constant temperature $T=T_0$ (i.e., constant $e=e_0=T_0/(\gamma-1)$). The isothermal hydrostatic, unmagnetized profiles, eq.~(\ref{eq:eos_rescaled}), then become
\begin{eqnarray}
    \hat T & = & 1~, \label{eq:background} \\
    \hat p_0(\hat z) & = & p_b e^{-\hat z} ~,\\
    \hat\rho_0(\hat z) &=& p_b e^{-\hat z} ~,\\
    \hat v_0^i & = & 0~, \\
    \hat B_0^i & = & 0~. \label{eq:background5}
\end{eqnarray}
Since the time evolution of $\hat\rho_0$ is zero, the continuity equation (\ref{eq:mhdnorm}) becomes
\begin{equation}
\partial_{\hat{t}}\hat{\rho}_1 + \partial_{\hat{k}} \left[ (\hat{\rho}_0 + \hat\rho_1) \hat{v}^k \right] = 0 ~, \label{eq:mhdnorm_pert}
\end{equation}
which allows to numerically evolve much more precisely the perturbations over a largely-varying range of densities. As a matter of fact, this approach generally avoids the numerical problems arising from inaccurate numerical operations due to the large numerical differences in the background fields among the extremes of the domain.

The perturbed pressure is calculated as
\begin{eqnarray}
  \hat p_1 &=& (\gamma - 1)\left(\hat E - \frac{\hat S^2}{2\hat \rho} - \frac{\hat B^2}{2}\right) - \hat p_0~, \label{eq:p1_def}
\end{eqnarray}
which implies, via the EoS:
\begin{eqnarray}
  \hat T_1 &=& (\gamma - 1)\frac{\hat E -\frac{\hat S^2}{2\hat \rho} - \frac{\hat B^2}{2}}{\hat \rho}  - \hat T_0 ~,\\
  \hat e_1 &=& \frac{\hat E -\frac{\hat S^2}{2\hat \rho} - \frac{\hat B^2}{2}}{\hat \rho}  - \hat e_0\,.
\end{eqnarray}

\subsection{Newtonian cooling}

In order to prevent the internal energy (hence, temperature and speed of sound) from rising indefinitely, we introduce an effective Newtonian cooling term, similarly e.g. to \cite{rogers14a}.\footnote{In the anelastic approximation, for which the temperature is usually evolved, the term usually enters in this way: $\partial_t T + ... = ... - T_1/\tau_{\rm cool}$, with $\nu_{\rm cool}=[(\gamma-1)\tau_{\rm cool}]^{-1}$.} The term enters in the right-hand side of the energy equation (\ref{eq:mhdnorm4}) as an additional source:
\begin{equation}
  \partial_{\hat t} \hat E + [...]= ... - \nu_{\rm cool} ~\hat p_1
\end{equation}
where we have used eqs.~(\ref{eq:eos_rescaled}) and (\ref{eq:p1_def}), and the parameter $\nu_{\rm cool}$ is the inverse of the cooling timescale. The latter is tuned to be short enough to remove excessive heating, but long enough to keep some finite perturbations in the internal energy (i.e., temperature, pressure). In our simulations, we set a value $\nu_{\rm cool} = 0.1$, an optimal choice for maintaining hydrostatic stability in our specific setup and method (see Appendix \ref{app:stability_tests}).

\section{Numerical setup}\label{sec:setup}

\subsection{Domain and boundary conditions}\label{sec:domain}

We assume a uniform background temperature, $T_0$, apt for the outermost layers ($p < 10$ bar, well above the RCB) of a highly irradiated planet \citep{youdin10}. Therefore, the background structure is given by eqs.~(\ref{eq:background})-(\ref{eq:background5}).

We set a Cartesian domain with $\hat{x}, \hat{y} \in [0,L/2]$, and the vertical direction $\hat{z} \in [0, L]$.
We choose $L=10$, so that $p_0$ and $\rho_0$ vary by a factor $e^{-10}=4.54\times 10^{-5}$. We consider a bottom pressure of 10 bar (i.e., $p_b=10$, using $p_{\rm bar}=1$ as code units), so that the pressure at the top of the domain is $\sim 0.45$ mbar. Taking the fiducial values appearing in eq. (\ref{eq:def_hp}) for $T_0$, $\mu$ and $g$, the chosen domain is ${\cal O}(10)\%$ of the typical radius of a HJ. The horizontal domain corresponds to $\lesssim$ 1 degree in latitude and azimuth, for typical values of planetary radius and $H_*$. Such domain hence represents a narrow column, and we will in particular consider steep wind profiles from GCM, corresponding to a region close to the dayside substellar point.

The boundary conditions in ${\hat x}$ and ${\hat y}$ are periodic. The vertical boundary conditions are less trivial. They are designed to: (i) ensure hydrostatic stability of the static background, which is delicate especially at the top; (ii) let the upwards perturbations to propagate in the damping region, where they are dissipated or go out of the domain, without any reflection. For all perturbed fields $(\hat \rho_1, \hat v^i, \hat p_1, \hat B^i)$, we impose a symmetric boundary condition over the ghost cells at the top and bottom layers, e.g., $\hat \rho_1(L)=\hat \rho_1(L-dz)$ (where $dz$ is the spatial spacing), meaning that $\partial_z \hat\rho_1=0$ among the two last cells. The only exception is for $\hat v_z$, for which we impose instead a reflective boundary condition among the last two numerical points, i.e. $\hat v_z(L)=-\hat v_z(L-dz)$.\footnote{In the Simflowny user interface, they are called {\em Reflection positive} and {\em Reflection negative}, respectively.} The boundary conditions on $(\hat\rho,\hat S^i,\hat E)$ are consistently imposed automatically from the perturbed and background fields, using the previous relations.

In the upper part of the domain, $\hat{z}>z_d$, we introduce a damping region, where all velocities are forced to go to zero. This helps to absorb any outgoing waves. It is enforced by adding a source in the momentum eq.~(\ref{eq:mhdnorm2}):
\begin{equation}
    {\cal S}_{d} = - A_d \frac{(\hat{z}-z_d)^2}{(L-z_d)^2} S_i~, \label{eq:damping}
\end{equation}
i.e. a gradually increasing damping factor, where we use $A_d=10$ and $z_d=8$ (see Appendix \ref{app:stability_tests} for a justification). 

The upper half of our domain presents therefore an artificial setup: (i) the damping region for $z_d>8$ just described; (ii) a wind intensity that unphysically decreases upward, to zero, as described below (\S~\ref{sec:forcing}). While these are unphysical features, we have verified that the amplitude of the magnetic field and currents generated here is much less than in the physical part, and, more importantly, the fluid dynamics propagate mostly upwards. No sign of artificially induced dynamics (e.g., spurious waves) is seen to appear at the top and propagate to the bottom, at least for the time over which we run most simulations ($t\sim 3000 ~t_*$).

We will focus mostly on the analysis below $\hat{z} < 5$, the physical region (leaving shaded in gray the artificial ones $\hat{z}\gtrsim 5$ in all plots hereafter). In particular, we will focus on the volume-integrated energies and on the plane-averaged and time-averaged values of the vector components. All the quantities are given in terms of the reference units defined above, by which they can be translated into physical units.

In Appendix \ref{app:stability_tests} we provide more information about how we have tuned the size of the domain, the damping region parameters and the boundary conditions, by assessing the stability of the background hydrostatic profile alone.

Placing the local domain and fields within a global view, $\hat z$ represents the radial direction, the $\hat x$ is the azimuthal direction, and $\hat y$ would be the latitude. Therefore, the $\hat y$-$\hat z$ components are a representation of the poloidal fields, and the $\hat x$ component represents the toroidal field.\footnote{In the poloidal-toroidal decomposition of a field, approximating the components as toroidal=azimuthal and poloidal=meridional plane is exact only in axial symmetry, although it is commonly done for simplicity.}

\subsection{Infrastructure and numerical methods}

We use v3.7.3 of Simflowny \citep{2013CoPhC.184.2321A,2018CoPhC.229..170A,palenzuela21}, a user-friendly platform that generates codes for partial differential equations, allowing a variety of numerical schemes in finite difference and finite volumes. It employs the SAMRAI\footnote{\url{https://computing.llnl.gov/projects/samrai}} infrastructure \citep{hornung02,gunney16} for the management of the parallelization, mesh refinement (here not used) and output writing. We employ the high-resolution shock-capturing method MP5 for the spatial discretization, using a splitting flux scheme, and a Runge-Kutta fourth-order scheme for the time advance\citep{palenzuela18}.

If the maximum velocities of the system are around the background speed of sound $c_{s,0}=\sqrt{\gamma}c_*$, then the Courant timestep is
\begin{equation}
    dt_c \lesssim \frac{dz}{\hat c_{s,0}} = \frac{L}{\sqrt{\gamma}N_z}~,
\end{equation}
where $N_z=L/dz$ is the number of points in the vertical direction. In particular, for $\gamma=7/5$ and $L=10$, our RK4 scheme implies a maximum timestep $dt_{\rm max} = 0.4~dt_c \lesssim (2.11/N_z)$. The vertical crossing time of the speed of sound is $t_z(L)\sim 8.45~t_* \sim 1$ hour.

We set $N_z=100$ in all the simulations shown here, with the same spacing in the three direction, i.e. $N_x=N_y=N_z/2$. We typically run our simulations for several thousands $t_*$, that would correspond to days (by chance, of the same order of magnitude of the computational clock time using a few tens of processors in our local cluster). However, time units are pretty unimportant, since our simulations aim at finding a stationary state, rather than evolving the system.

Since our numerical scheme does not conserve by construction the Maxwell divergence constraint $\vec{\nabla}\cdot\vec{B}=0$, a divergence cleaning scheme \citep{dedner02} is implemented like in \cite{palenzuela18}, to which we refer also for further details and references about the numerical scheme and benchmark tests of the code. The constraint is well maintained during the entire evolution in our simulations.

\subsection{Forcing: zonal wind vertical profile}\label{sec:forcing}

Transonic winds of up to several km/s at a pressure of fractions of bar are predicted by all GCMs (e.g. \citealt{showman02,showman10,heng11}), and are indeed inferred from radial velocity spectral shifts (e.g. \citealt{snellen10}) and by an observationally inferred eastward advection of the hottest point in HD 189733b \citep{knutson07,knutson12}. Existing models predict $\sim$km/s winds down to $\sim 1-10$ bar in the equatorial regions. Slower flows in the opposite direction are usually present at larger depths. Nevertheless, the opposite trend has been proposed, with outer retrograde equatorial flows and inner superrotational flows \citep{carone20}.
Wind speed can be slowed down by the magnetic drag \citep{perna10a,rogers14a}. In any case the vertical profile depends on the longitude and on the temperature.

\begin{figure}
\centering	\includegraphics[width=0.9\linewidth]{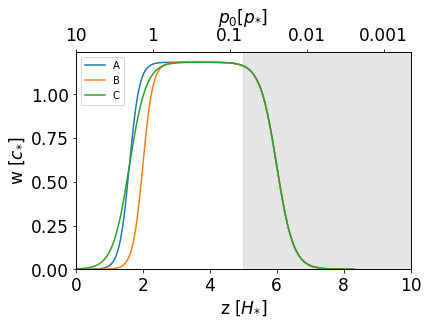}
\caption{Wind profiles adopted in this work, as a function of height ${\hat z}$, and background pressure $p_0$ (in units $p_*$, which we take 1 bar by default), for the models A, B, C of Table~\ref{tab:models}. The damping region (see text) is shaded in grey.}
\label{wind_profiles}
\end{figure}

As an effective approach, we here assume a zonal wind that mimics the main features seen in GCMs (e.g. Fig. 4 of \cite{beltz22} or right panel of Fig. 2 of \cite{rogers14a}), using the least possible parameters. For simplicity, we consider a zero zonal velocity at the bottom of the domain, since at those pressure (10 bar) the zonal velocity is one or two orders of magnitude smaller than at the top.\footnote{Seen in another way, we choose the frame which is co-moving with the wind at the bottom of the layer; results are insensitive to the frame choice.} Additionally, in order to ensure stability in the top part, we prescribe a transition to zero velocity in the damping region. This creates a second, artificial shear layer, but it doesn't affect the results in the physical domain of the box, as said above. Considering this, the implemented wind profile reads:
\begin{equation}
w(z) = w_0\sqrt{\gamma}\left(\frac{1+{\cal B}}{2}\right)\left(\frac{1+{\cal T}}{2}\right)~, \label{eq:vwind}
\end{equation}
where
\begin{eqnarray}
    {\cal B}&=&\tanh\left(\frac{{\hat z}-z_b}{\alpha_{b}}\right) \\
    {\cal T}&=&\tanh\left(\frac{z_t-\hat{z}}{\alpha_{t}}\right)~,
\end{eqnarray}
and $z_t,z_b$ correspond to the height of the shear layers at the top (which is fictitious) and at the bottom (the real one), respectively, and {$\alpha_t$}, {$\alpha_{b}$} the corresponding shear layer thickness. This implies that the middle region ($\hat z \sim$ ($z_{b}$+$\alpha_{b}$, $z_t$-$\alpha_{t}$)) has a characteristic speed $w=\sqrt{\gamma}$, i.e. the speed of sound. Fig.~\ref{wind_profiles} displays the different wind profiles that have been used in this project. We consider different parameters for the physical shear layer: depth, taking $z_{b}$=1.6 or $z_{b}$=2, and thickness, taking $\alpha_{b}$=0.3 or $\alpha_{b}$=0.5. We fix $z_t=6$ and $\alpha_{t}$=0.5, a choice that, within a preliminary exploration of different values (in combination with the other numerical parameters as discussed in \ref{app:stability_tests}), allows stability and minimize the influence of the artificial shear layer on the rest of the domain.

The effective forcing is defined as
\begin{equation}
\hat{F}_{\rm src} = \hat F_0 \hat \rho (\max\{w({\hat z}) - \ \langle v_x \rangle, 0\}\vec{e}_{x} + \vec{\delta v}(\vec{x}))~,
\label{eq:forcing}
\end{equation}
where: $\langle v_x \rangle$ is a local average of $v_x$ in the $x$-direction (across a few points); $\vec{\delta v}$ are random perturbations (see below); $\hat F_0:=F_0/g$ sets the inverse of the timescale over which the velocity responds to the forcing (set to 1 here); hereafter $\vec{e}_i$ indicates the unit vector along a direction $i$. Including this average is not strictly needed, and is only motivated by what done in \cite{ryu18}. As an alternative, we also try the same, but with $\langle v_x \rangle=v_x$, i.e., without the average.  We will compare below the results coming from the different prescriptions (see e.g. model A vs model AS in Table \ref{tab:models}).

We prescribe the initial value of $v_x=w(z)$, and, with this forcing prescription (in either flavours), we keep $v_x$ close to the wind profile (compensating the numerical dissipation). However, this choice lets $v_x$ slowly vary in time and become locally slightly larger than $w(z)$, allowing positive deviations from the simple wind profile. Such deviations provide winds that slowly vary in time and lead to an increase of the total kinetic energy (see below). Although this arbitrary choice can be improved, this doesn't hamper the main conclusions.

Notice that since we perform a local simulation with a prescribed external forcing, there is no effective feedback on the velocities from the magnetic drag.

\subsection{Forcing: turbulent perturbations}

The additional perturbations of eq.~(\ref{eq:forcing}) represent local deviations from the mainly zonal flow $w(z)$, as:
\begin{equation}
 \vec{\delta v} = \lambda~(r_1 \vec{e}_{x} + r_2  \vec{e}_{y} + r_3 \vec{e}_{z})~w(z)~,
\end{equation}
where $\lambda$ is a constant parameter regulating the average relative amplitude $\delta v(z)/w$, and $r_{1}$, $r_{2}$ and $r_{3}$ are random numbers $\in [-1,1]$ associated to each direction. The random variations are totally uncorrelated from point to point and from one timestep to another (meaning that, effectively, the perturbation are correlated over a time $dt$).
 
The introduction of the turbulent contribution to the forcing throughout the domain is a simple, effective way to consider at the same time the intrinsic time variability of the wind \citep{menou20}, and the injection of small scale perturbations. Such variability and perturbations are observed in the outer atmosphere of Jupiter and are arguably present in Hot Jupiters, as past studied considered \citep{li10,youdin10,menou22}. The origin of such variability lies in the intrinsic non-linearity of the problem, and can be further fed by e.g. the presence of secular shear layers, latitude and longitude dependence of the irradiation and local inhomogeneities in the chemical composition.

The necessary condition for developing turbulence from a shear layer is that the Richardson number (ratio between buoyancy and flow shear terms) is below a critical value,
\begin{equation}\label{eq:richardson}
    {\cal R}i= \frac{g |\partial_z(\ln\rho)|}{(\partial_z v_x)^2} < 1/4~.
\end{equation}
This critical value can be a bit higher (therefore, making turbulence easier) when thermal diffusivity is taken into account, as shown by \cite{li10}. However, it is important that the shear layer is smaller than the scale height. In our case, $\hat g=\partial_z \ln\hat\rho_0=1$ from eq.~(\ref{eq:vwind}), so that $\partial_z \hat v_x \sim$ $\alpha_{b}^{-1}$ $\gtrsim 2$. We therefore test values $\alpha_{b}\lesssim 0.5$, indeed in line with the GCM wind profiles discussed above.

Note that, in other studies, Kelvin-Helmholtz box simulations apply only an initial perturbation to $\vec{v}$ \citep{ryu18,vigano19}, and study the properties of the slowly decaying turbulence. We choose instead to apply a continuous small-scale forcing, since we seek to achieve a stationary state (not a decaying one), sustained by the persistent climate-induced winds (including the small-scale variations) acting on our small column. Moreover, for the relatively low resolution used here (see below), imposing only an initial perturbation would hamper or quench rapidly the turbulence.

\subsection{Initial magnetic field}\label{sec:initial_b}

Under the presence of a large-scale vertical component of the magnetic field (generated in the interior), the most relevant effect is its winding. As a matter of fact, if for simplicity we consider a wind profile $v_x(z)$ and a purely uniform magnetic field along any direction, the ideal induction equation reads
\begin{equation}
    \frac{\partial \vec{B}}{\partial t}= \vec{\nabla }\times \left(\vec{v} \times \vec{B} \right)= B_z\frac{\partial v_{x}}{\partial z} \vec{e_{x}}\,,
\end{equation}
since $\vec{v} \simeq w(z) \vec{e}_x$. This means that, provided a non-null value of $B_z$, there is at the beginning a continuous generation of the azimuthal ($x$) component of magnetic field, linear in time:
\begin{equation}
    B_{x}= B_{z} \frac{\partial v_{x}}{\partial z}t~,
    \label{diveBintime}
\end{equation}
which shows how the horizontal field grows in correspondence to the shear layers of the wind. This linear growth saturates when either the resistivity starts counter-acting the field growth, or when the magnetic drag becomes important. In our case, however, the shear flow is time-independent due to the forcing, therefore the magnetic drag does not act. The only force stopping an indefinite growth of the field is then the numerical resistivity, which depends on the resolution (see \S 
4.1).
Since we do not have an explicit physical resistivity implemented, we initialize the magnetic field with an equipartition guess for the saturated field $B_x^2 \simeq \rho v_x^2$. Such guess should not be far from reality if we are considering the physically realistic values of the wind and the numerical resistivity (i.e., the resolution). Using such a prescription, together with the chosen profile $\frac{\partial v_{x}}{\partial z} \sim w$/$\alpha_{b}$, gives a saturation timescale (marking the end of the linear growth) of:
\begin{equation}
t=\sqrt{\rho}\frac{\alpha_{b}}{|B_{z}|}\,,
\end{equation}
that we use in eq.~(\ref{diveBintime}) to set the initial magnetic field:
\begin{eqnarray}
    B^{in}_{x}(z) &=& C_B~{\rm sgn}(B^{in}_z)~\sqrt{\rho_0(z)}~\frac{\partial w}{\partial z}(z)~\alpha_{b} =   \label{eq:initial_b}\\
    & = & C_B~{\rm sgn}(B^{in}_z)~\sqrt{\rho_0}~w\left(1-{\cal B} - \frac{\alpha_{b}}{\alpha_{t}}(1-{\cal T}) \right)~,\nonumber
\end{eqnarray}
where we have employed the wind velocity prescription eq.~(\ref{eq:vwind}). $C_B$ is a parameter of order one that we vary to test different initial amplitudes, since our guess is not rigorous.

Additionally, we consider an initial small, uniform magnetic field in the vertical direction, $B^{in}_z$, representing the one generated inside. The initial field $B^{in}_x$, eq. (\ref{eq:initial_b}), only depends on the direction of the planetary field, which sets the direction of the winding. Such induced field $B_x(z)$ is sustained by currents $J_y(z)$, which, as we will see, represent the dominant component. If we consider this local model part of a global, spherical one, these would be the latitudinal currents. They enclose a toroidal field in a this shearing shell, which can be extremely intense (see below), but completely screened outside.

Starting from the initial magnetic field, eq. (\ref{eq:initial_b}), allows us to save computational time during the linear growth, entering directly in the non-linear dynamics. In particular, we start with $C_B=1$, since lower or higher values imply a much longer time to approach the asymptotic winding-generated profile of $B_x$, as shown in the 1D tests (i.e., $\lambda=0$) in Appendix~\ref{app:initial_b}.

\section{Results}\label{sec:results}

We have explored different initial configurations, in order to study the general behaviour and to quantitatively assess the magnetic field and current profiles. In Table~\ref{tab:models} we present the models with the different initial parameters which are varied (varying only one compared to the reference model A). In particular, we study the dependence on the wind profiles shown in Fig.~\ref{wind_profiles} (compare models A, B, C), and study the role of some forcing details: the perturbation amplitude (A0, A, Al, Ah), time correlation (Adt), and with/without the average over $v_x$ (without for AS, AlS, AhS).  In all simulations, we fix the same values of the numerical parameters ($\nu_{\rm cool}$, $A_d$, $z_d$, as discussed above) and for the other physical parameters: $\gamma=1.4$, $C_B=1$, ${\hat B}_z(t=0)=0.0001$, ${\hat B}_y(t=0)=0$, $w_0=1$.

\subsection{The role of resolution and numerical resistivity}

\begin{figure}
\centering
\includegraphics[width=.9\linewidth]{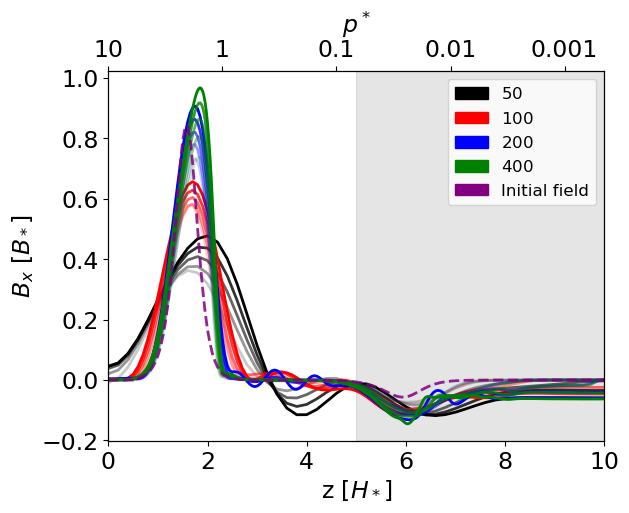}\\
\includegraphics[width=.9\linewidth]{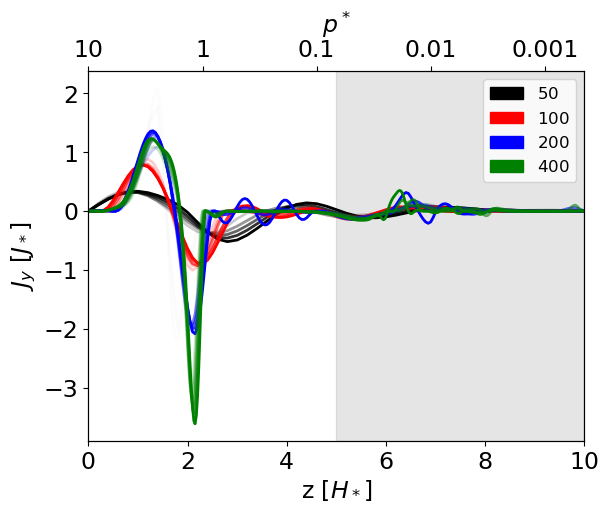}
\caption{Comparison of the vertical profile evolution of $B_x$ (top) and $J_y$ (bottom) in the 1D problem, i.e., in the absence of perturbations, $\lambda=0$ (model A0). The four colors indicate different values of the vertical resolution used, $N_z$ (legend). The five shades (transparent to opaque) indicate different times: $t/t_*=500,1000,1500,2000,2500$. For the top figure the initial magnetic field, i.e. $t/t_*=0$, common for the four resolutions, has been plotted in a purple dashed line.} In 1D, results converge rapidly since no perturbations cause stochastic variations.
\label{fig:resolution}
\end{figure}

The results presented here are valid as long as the physical magnetic diffusivity $\eta=1/(\sigma \mu_0)$, neglected in this work, is not much larger than the numerical one, $\eta_{\rm num}$. If not, we underestimate the effect of the physical resistivity that limits the winding growth, and thus we overestimate the magnetic fields.
Indeed, as a first indication, Fig.~\ref{fig:resolution} shows the dependence of the winding-controlled shape of $B_x$ and $J_y$ on the resolution, for 1D models (i.e., with $\delta=0$, so no $x$ or $y$ dependence is included). As we increase the resolution, the magnetic field gets larger and larger, with a difference in peak of a factor $\sim 2$ between $N_z=50$ and 400. This is because increasing the resolution means having less numerical resistivity, which is the only term counter-acting the winding (recall that magnetic drag is not active due to the fixed velocity forcing). Therefore, this lack of numerical convergence is expected. Moreover, note that the order of magnitude of the quantities is the same (hundreds of G for the maximum value of $B_x$). The natural question is then which resolution we should adopt. In App. \ref{app:diffusivity} we estimate the numerical resistivity and we show how very hot Jupiters ($\gtrsim 3000$ K) have physical resistivities of the order of the numerical diffusivity with $N_z\sim 100$, which is the resolution we adopt hereafter. Such a resolution is enough to resolve the shear layer and to assess the turbulent motions. Higher resolutions would need the inclusion of the physical magnetic diffusivity.

We reserve the implementation of a realistic profile of the physical magnetic diffusivity, as well as higher resolutions, to a follow-up work.

\begin{table}
\centering
\caption{List of the parameters varied in the different models. In the first column we indicate the name of the simulation where A is the default simulation; B, C explore modifications of the wind profile (see the height $z_b$ and thickness $\alpha_{b}$ of the shear layer, second and third columns); l and h refer to lower or higher values of perturbations $\lambda$ (fourth column); dt refers to a change in the timestep (i.e., the time-correlation between random perturbations, fifth column); S to take the simple local velocity $v_x$ without the local average in the forcing, eq.~(\ref{eq:forcing}). The sixth column indicates the numerical integral ${\cal I}:=\int_0^5 \hat{J}^2 \exp{(-\hat{z}/2)}~d\hat{z}$, averaged in the $x-y$ plane and in time (i.e., over all the tens of 3D outputs available after $t\gtrsim 500~t_*$), eq.~(\ref{eq:int_I}). Such quantity is related to the a-posteriori Ohmic dissipation estimation, see \S\ref{sec:applicability}. The last column indicates the maximum value of the temporal and spatial average of $B_x(z)$ in kG (taking $p_{\rm bar}=1$, i.e., a pressure at the bottom of $p_b=10$ bar).}
\label{tab:models}
\begin{tabular}{|c|c|c|c|c|c|c|}
\hline Name & $z_{b}$&$\alpha_{b}$ &${\lambda}$ & $dt~[t_*]$ & ${\cal I}$ & {\rm max}$(\langle B_x \rangle)[kG]$\\
\hline A & 1.6 & 0.3 & 0.01 & 0.0025 & 0.26 &5.3 \\
\hline B & 2 & 0.3 & 0.01 & 0.0025 & 0.096 &5.6 \\
\hline C & 1.6 & 0.5 & 0.01 & 0.0025 & 0.46 &5.1\\
\hline A0 & 1.6 & 0.3 & 0 & 0.0025 & 0.36 &6.1\\
\hline Al & 1.6 & 0.3 & 0.001 & 0.0025 & 0.30 &5.6\\
\hline Ah & 1.6 & 0.3 & 0.1 & 0.0025 & 0.28 &5.9\\
\hline Adt & 1.6 & 0.3 & 0.01 & 0.01 & 0.18 &1.5\\
\hline AS & 1.6 & 0.3 & 0.01 & 0.0025 & 0.30&5.4\\
\hline AlS & 1.6 & 0.3 & 0.001 & 0.0025 & 0.31&5.4\\
\hline AhS & 1.6 & 0.3 & 0.1 & 0.0025 & 0.24&5.3\\
\hline
\end{tabular}
\end{table}

\subsection{General behaviour}

We begin with a detailed analysis of the reference simulation A, which we ran for a very long time ($6000 ~t_*$), having $\lambda=0.01$, $\alpha_{b}$=0.3, ${\hat z}_b=1.6$, $C_B=1$, a conservative timestep $dt=0.0025~t_*$, and employs a local average $\langle v_x \rangle$ in the forcing, eq.~(\ref{eq:forcing}). The general behaviour described here is common to all other simulations that we have analyzed, though we follow them for shorter times, $t \sim 2000-3000 ~ t_*$.

\begin{figure*}
\centering
     \subfigure[]{%
        \includegraphics[width=0.2091\linewidth, height=0.24\textwidth]{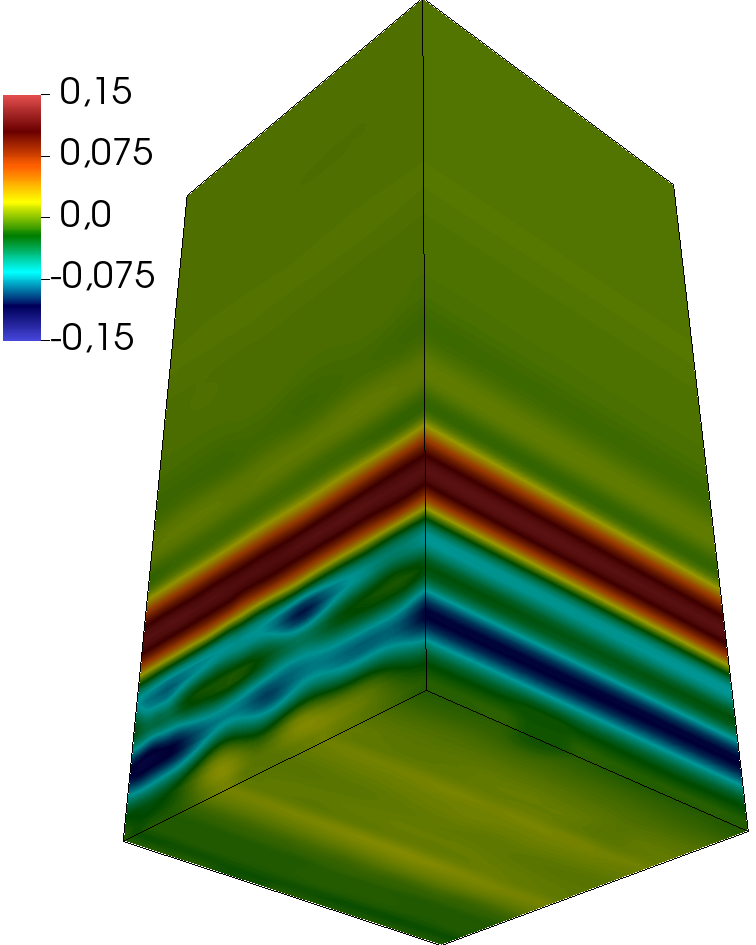}
        \label{fig:1}}
     \subfigure[]{
        \includegraphics[width=0.2091\linewidth, height=0.24\textwidth]{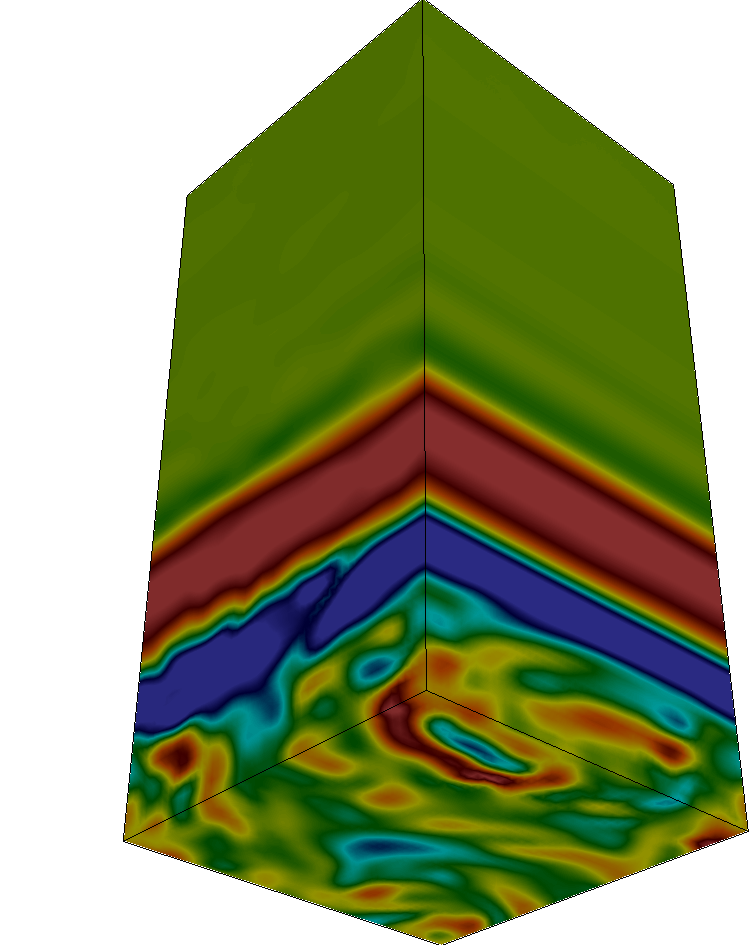}
        \label{fig:2}}
         \subfigure[]{%
        \includegraphics[width=0.2091\linewidth, height=0.24\textwidth]{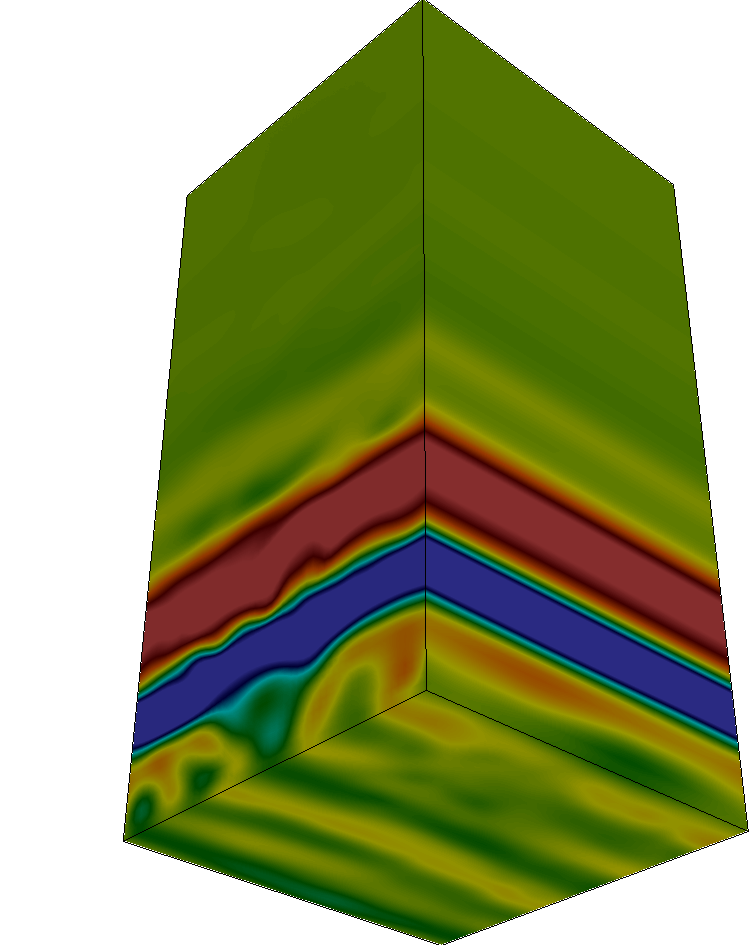}}
        \label{fig:3}
     \subfigure[]{
        \includegraphics[width=0.2091\linewidth, height=0.24\textwidth]{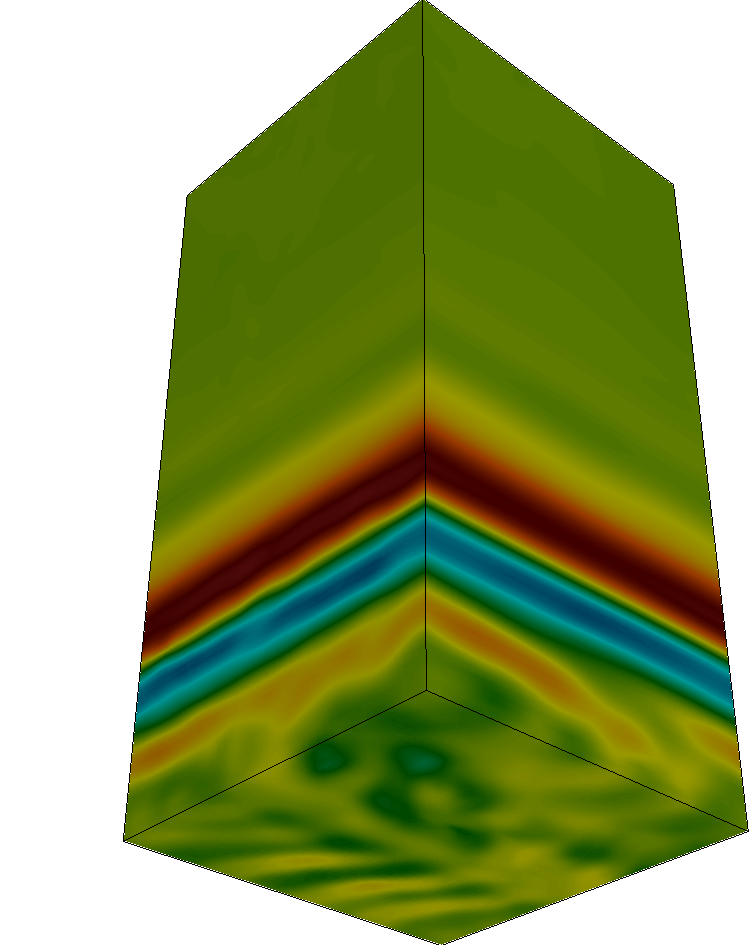}
        \label{fig:4}}
\caption{Vertical 3D snapshots of the perturbed density $\rho_1$ (scale color, in units $\rho_*$) for simulation A, at four different times $t/t_*$=500 (a), 2000 (b), 3000 (c) and 6000 (d). The shear layer is centered at $z_b=1.6$.  The values related to the color scale are in units of $\rho_*$.} 
\label{rho1}
\end{figure*}

In Fig.~\ref{rho1} we show a 3D box of the density perturbation $\rho_1$, for four different timesteps of the simulation between $t/t_*$=500 and 6000. Such perturbations, generated by the turbulent state that quickly develops, are typically of the relative amplitude $\rho_1/\rho_0\sim {\cal O}(0.01)$. The perturbations $\rho_1$ (and the corresponding $p_1$, not shown here), depend mostly on $z$. However, in the lower part of the domain some 3D structures appear due to the random perturbation enforced. These structures show a pseudo-oscillatory trend of increasing (e.g., second panel) and decreasing amplitudes (by a factor of a few) on timescales of roughly a thousand $t_*$. In the upper unphysical region there are no significant perturbations; only at very late times $t\gtrsim 4000~t_*$, some perturbations pile up, but in any case they do not propagate downwards. Such piling up is probably due to the magnetic field easily induced in the uppermost regions, despite the damping in the velocities.

\begin{figure*}
    \centering
     \subfigure[]{%
        \includegraphics[width=0.2091\linewidth, height=0.24\textwidth]{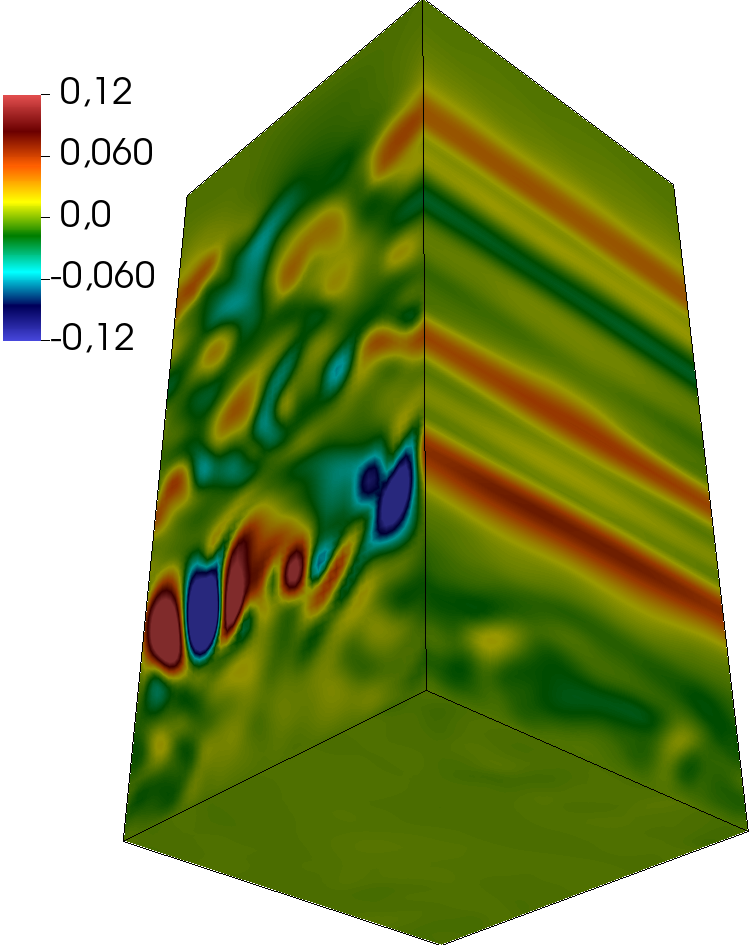}
        \label{vz}}
    \subfigure[]{%
        \includegraphics[width=0.2091\linewidth, height=0.24\textwidth]{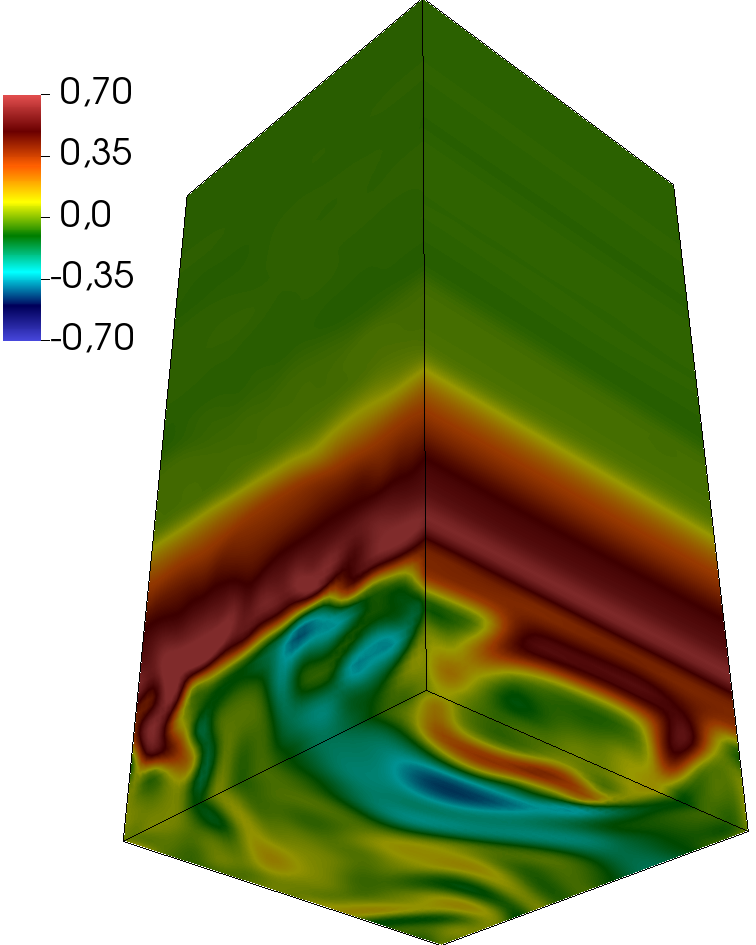}
        \label{fig:bx3d}}
     \subfigure[]{
        \includegraphics[width=0.2091\linewidth, height=0.24\textwidth]{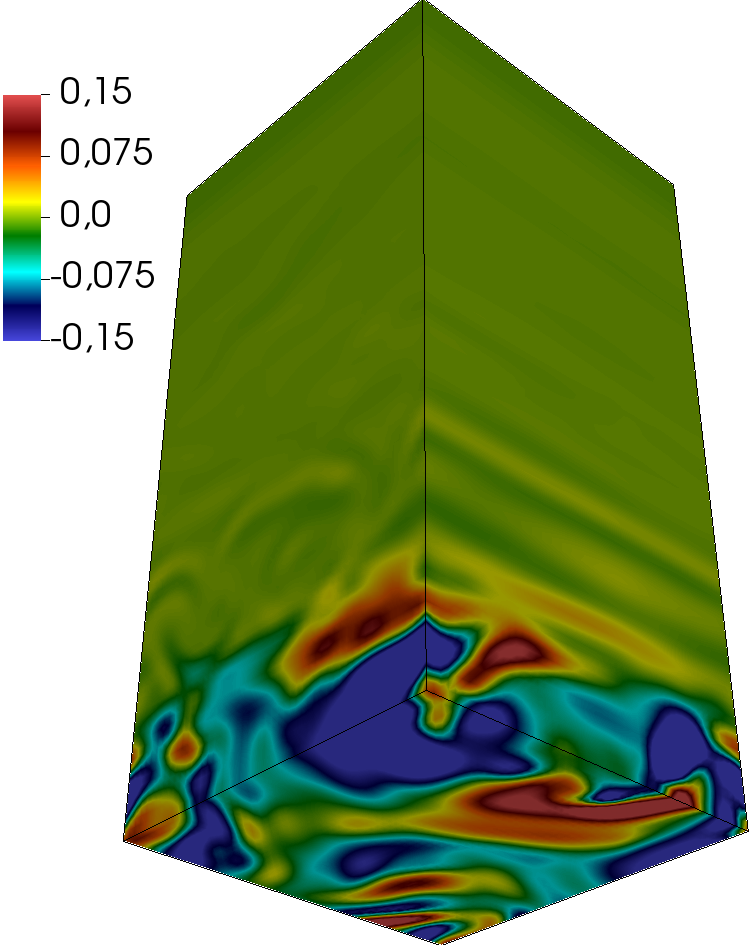}
        \label{fig:by3d}}
     \subfigure[]{%
        \includegraphics[width=0.2091\linewidth, height=0.24\textwidth]{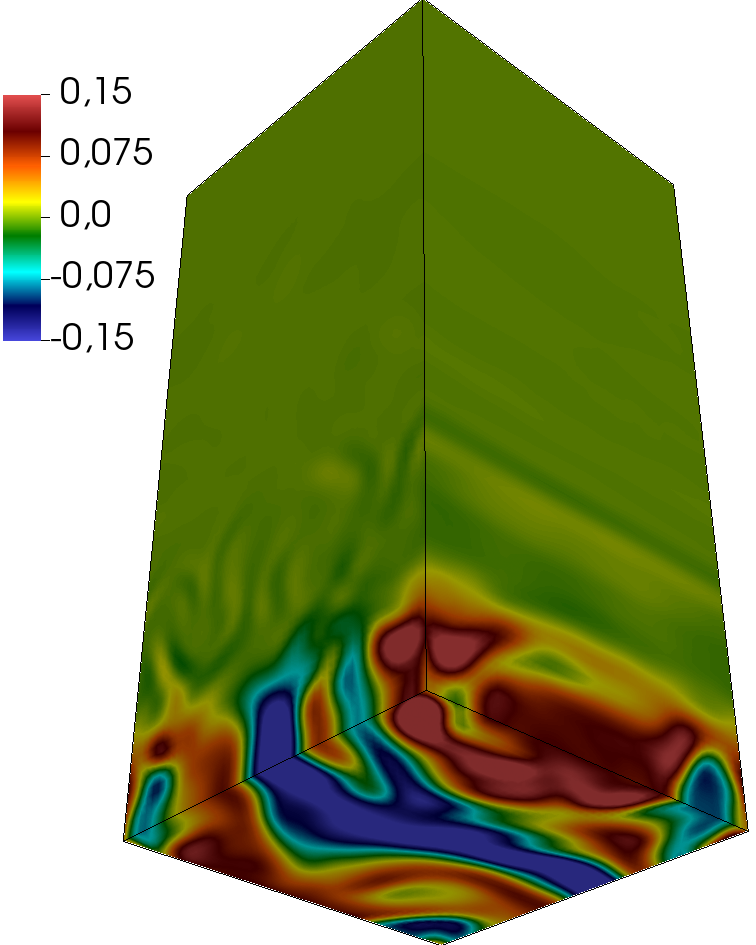}}
        \label{fig:bz3d}
\caption{Vertical 3D snapshots of $v_z [c_*]$ (a), $B_x [B_*]$ (b), $B_y [B_*]$ (c) and $B_z [B_*]$ (d), for simulation A, at $t=2000~t_*$. } 
\label{fig:3d_slices}
\end{figure*}

\begin{figure}
\centering
\includegraphics[width=0.9\linewidth]{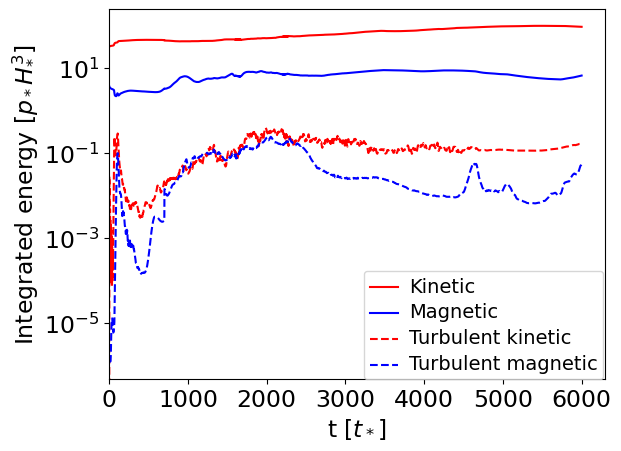}\\
\includegraphics[width=0.9\linewidth]{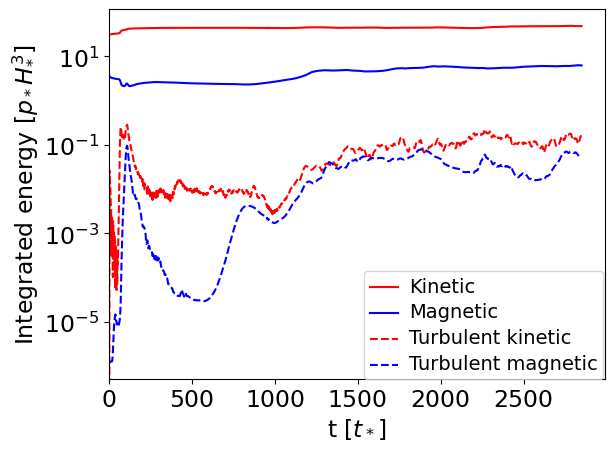}
\caption{Volume-integrated kinetic (red) and magnetic (blue) energies, for the representative simulations A (top), followed for a very long time, and AS (bottom). Solid lines represent the total value, overwhelmingly dominated by the $x$ component (zonal). Dashed lines represent the turbulent contributions ($y$ and $z$ components, i.e. meridional and vertical).}
\label{fig:integrals}
\end{figure}

The upward propagation is explicitly seen in the corresponding $v_z$ vertical 3D snapshot (first panel of Fig.~\ref{fig:3d_slices}): the fluid motion structure is dominated by positive values (red), although at mid latitude there are turbulent structures in the $y-z$ plane. The $x$-dependence is instead limited, due to the forcing. As for $\rho_1$, the $v_z$ keeps fluctuating around small values. In general, the vertical motion is the result of the following three contributions:
\begin{enumerate}
    \item The main one comes from the presence of the magnetic pressure in the shear layers (the physical one around $z_b=1.6$ and the artificial one around $z_t=6$), with values of a few percent of the speed of sound. At later times, the upward motions dominate but they are efficiently absorbed by the damping region at $\hat{z}>8$.
    \item The second contribution comes from the turbulent motions, leading to the 3D vertical structures seen in both $v_z$ and $\rho_1$ (especially at the time shown here).
    \item Finally, a third source of vertical motion comes from the hydrostatic readjustment due to discretisation and is seen even in non-magnetic 1D tests for stability (i.e., without perturbation, hence no $x$ or $y$ dependencies, see Appendix~\ref{app:stability_tests}). However, this latter numerical contribution is orders of magnitude smaller compared to the other two physical sources.
\end{enumerate}
Generally speaking, after a transitory phase from the initial conditions lasting $\lesssim 1000~t_*$ (a hundred crossing timescales), the system tends to reach a quasi-stationary state, where winding and numerical diffusivity reach a balance (see the 1D simulations with $\lambda=0$ in Appendix~\ref{app:initial_b}). This general behaviour is reflected by the kinetic and magnetic energies, integrated over the domain, as shown in Fig.~\ref{fig:integrals}, for models A and AS. Both the velocity and magnetic fields are dominated by the $x$ (zonal) components since the beginning. The total magnetic energy stays within a factor of a few of its initial value, thanks to the educated guess on $B_x$, eq.~(\ref{eq:initial_b}). The total kinetic energy experiences a slight secular increase, due to the forcing function we assume, eq.~(\ref{eq:forcing}) (see discussion in \S\ref{sec:forcing}). It is very useful to look at the turbulence-induced components, $y$ and $z$. The turbulent magnetic energy sharply increases in the very early times, then stabilises around $0.01-0.1~p_*H_*^3$. On the other hand, the turbulent kinetic energy shows less variations and is more stable. The AS model looks slightly more stable, especially in the total kinetic energy, due to the different prescription of the forcing.

\subsection{Magnetic fields topology}

\begin{figure}
\centering
\includegraphics[width=0.5\textwidth]{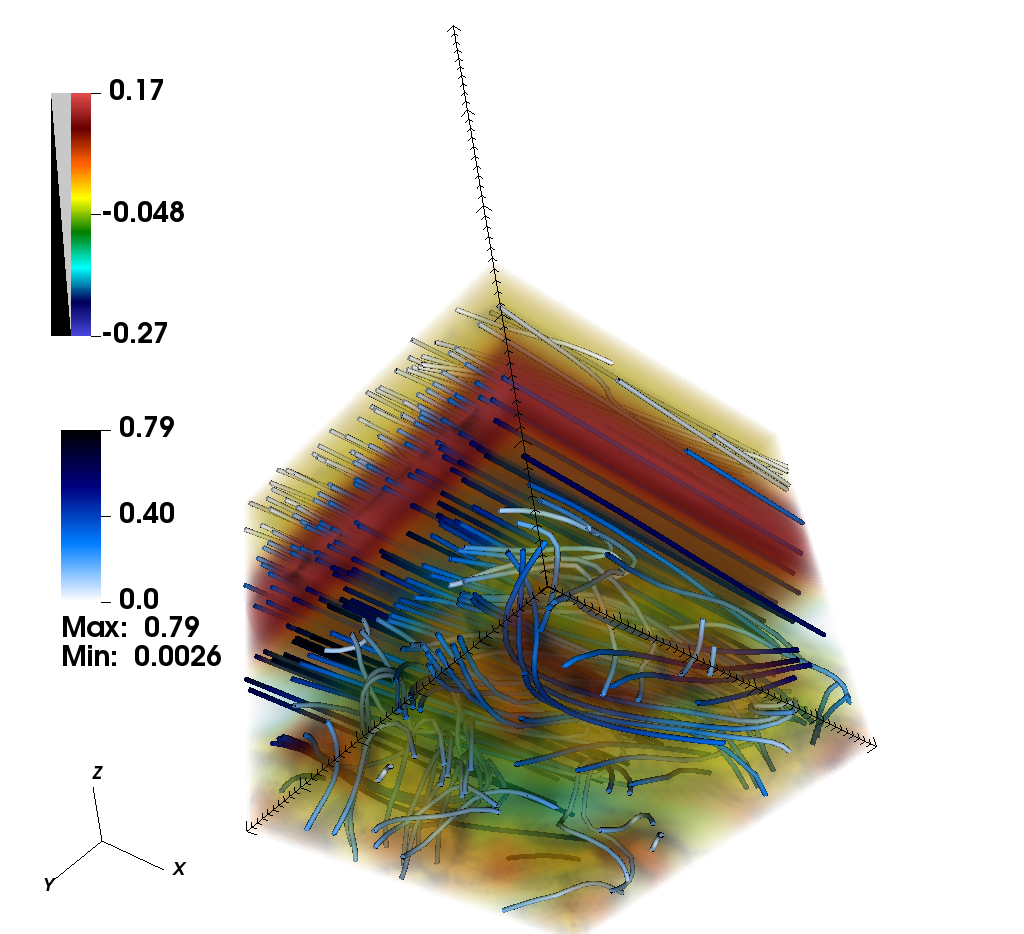}
\caption{Representative 3D snapshot of simulation A, at $t=2000~t_*$. The rainbow color scale represents the value of $\rho_1$ in units $\rho_*$, while the magnetic field lines are colored with the intensity of the magnetic field. We focus here only on the physical part of the domain, $\hat z < 5$.} 
\label{3D_results}
\end{figure}

We now focus on the properties of the magnetic fields at equilibrium. Fig.~\ref{3D_results} is a representative snapshot of the magnetic fields (lines in blueish scale) and $\hat \rho_1$ (rainbow-like colors) for simulation A, at $t=2000~t_*$.
Fig. \ref{fig:3d_slices}, second, third and forth panels, also shows the snapshots of $\hat B_x$, $\hat B_y$ and $\hat B_z$ at the same time. 
The strongest magnetic field in the $\hat x$ direction is generated in the shear layer. The magnetic field is clearly composed by structures are elongated in the $\hat x$ direction, due to the winding. However, they show a certain complexity in the $\hat y$-$\hat z$ plane, reflecting the curling and twisting effect of MHD turbulence. Such structures tend to periodically become more or less complex, similarly to what is seen for $\hat \rho_1$ (here we show a snapshot at $t=2000~t_*$ with a particularly rich topology, on average they are less visible).

\begin{figure}
\centering
\includegraphics[width=0.9\linewidth]{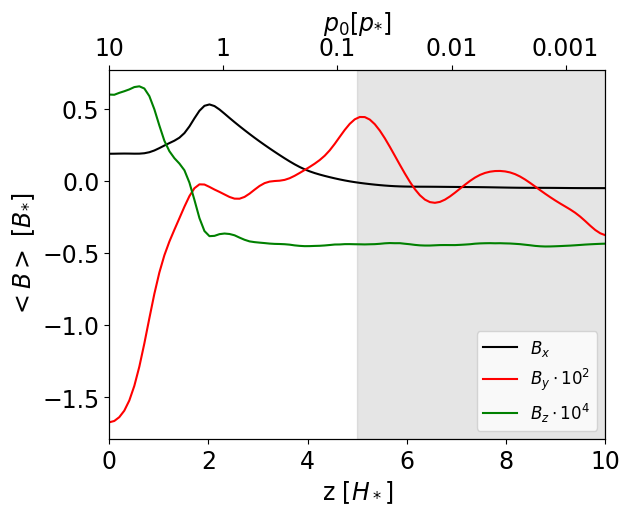}\\
\caption{The vertical profiles $\langle \hat B_{x}\rangle(\hat z)$, $\langle \hat B_{y}\rangle (\hat z)$ and $\langle \hat B_{z}\rangle(\hat z)$ averaged over the ${\hat x}$-${\hat y}$ plane. The averaged value for all the times after $t/t_*>500$ for all the components is shown. $B_{x}$ is in black, $B_{y}$ in red and $B_{z}$ in green. We indicate the background value of the pressure, $p_0$, in units of $p_*=1$ bar. The shaded area represents the unphysical region, which for numerical reasons includes an artificial shear and a damping layer.}
\label{fig:BxByBz}
\end{figure}

In Fig. \ref{fig:BxByBz}, we show in a clearer way the evolution of the three magnetic field components, averaged over the ${\hat x}$-${\hat y}$ plane. They represent time-averaged quantities for $t/t_*>500$. Hereafter, the shaded area in any plot represents the unphysical region (see \S\ref{sec:domain}), which anyway encloses a negligible fraction of the magnetic energy, currents and vorticity produced.

Let's start with the dominant component, $\langle\hat B_{x}\rangle(z)$. It readjusts from the initial profile, showing in particular an increase for $\hat{z} \in (2,4)$. The average has been calculated for values $t_{*}>500$ since after this time the profile keeps within the same order of magnitude with only some fluctuations due to the movement of the fluid, as the stationary equilibrium has been reached.
At saturation, the maximum value $B_x\sim 0.5 B_*$ is slightly above the shear layer, $\hat{z} \sim 2$  (see Table 1). The magnetic field is negligible for $\hat{z}\gtrsim 5$. These values corresponds to local intensities up to a few $\sim$ kG in the shear layer (in the physical units discussed in \S\ref{sec:rescaled_eqs}), while just outside that the field drops by one order of magnitude or more.

Note also that at the bottom of the the domain (${\hat z}<1$) there is a non-negligible value of $\hat B_x$. This is due mostly to the numerical diffusion, combined with the bottom boundary conditions that allow non-zero vales. The boundary conditions enforce zero tangential currents ($\hat J_x$, $\hat J_y$), while radial currents $\hat J_z$ can penetrate to deeper layers. This is interesting, since propagation to deeper levels has direct consequences for the inflation of the planetary radii. A more rigorous characterization of these details needs resistive MHD simulations, which will allow us to understand better what comes from physics and what is numerical.

Compared to $\langle\hat B_x\rangle$, the other two components (spatially averaged in the same way), $\langle\hat B_{y}\rangle$($\hat z$) (middle panel) and $\langle\hat B_{z}\rangle$($\hat z$) (bottom) are a few orders of magnitude smaller, with more fluctuations around zero. The fact that $\langle\hat B_{z}\rangle$ is substantially smaller than $\langle\hat B_{y}\rangle$ is a reflection of the stratification that makes fluid motions (and therefore, magnetic amplification by stretching) easier in the horizontal directions rather than vertical one. Importantly, the vertical field in the outer region has a pretty flat behaviour, with much smaller values, confirming that the atmospherically induced fields are screened outside the shear layer.

\subsection{Currents}

\begin{figure}
\centering
\includegraphics[width=0.9\linewidth]{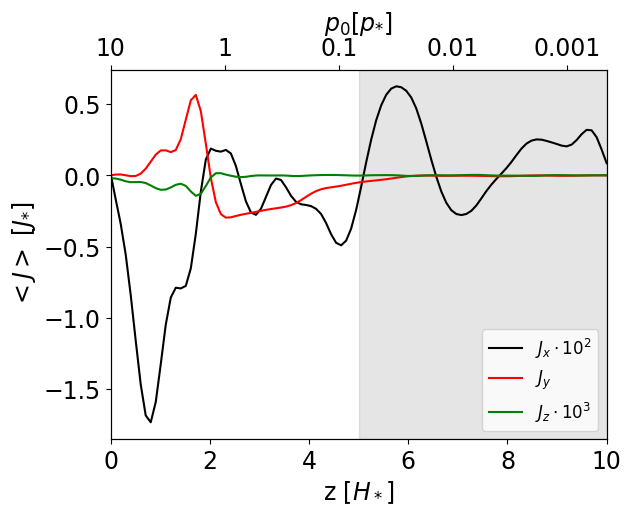}\\
\caption{Same as Fig.~\ref{fig:BxByBz}, but for $\langle \hat J_x\rangle(\hat z)$  (amplified by a factor of 100 for clarity)}, $\langle \hat J_y\rangle(\hat z)$  (amplified by a factor of 1000) and $\langle\hat J_z\rangle(\hat z)$.
\label{fig:JxJyJz}
\end{figure}

We now move to examine the sustaining atmospheric currents, showing the vertical profile, component by component, in Fig.~\ref{fig:JxJyJz}, again averaged over the ${\hat x}$-${\hat y}$ plane. Looking at the time-averages (again at times $\gtrsim 500~t_*$), we note that the dominant component is $\langle\hat J_y\rangle(z)$ (in the plot, $J_x$ and $J_z$ are amplified by factors of 100 and 1000, respectively). It has a maximum value at around $\hat z\sim 1.8$, close to the shear layer center $z_b$. There is a readjustment from the initial value (given by the $z$-derivative of eq.~\ref{eq:initial_b}). However, there is a change of sign across the shear layer, indicative of the two sides of the meridional loop of currents that enclose the generated toroidal field. The peak in the deep side (low ${\hat z}$) is sharper than the spatially extended negative currents in the shallow side (high ${\hat z}$).
Finally, the remaining two components oscillate instead around zero, with different amplitudes, reflecting again their stochastic turbulent behaviour.

\subsection{Dependence on the wind profile}

\begin{figure}
\centering
\includegraphics[width=1\linewidth]{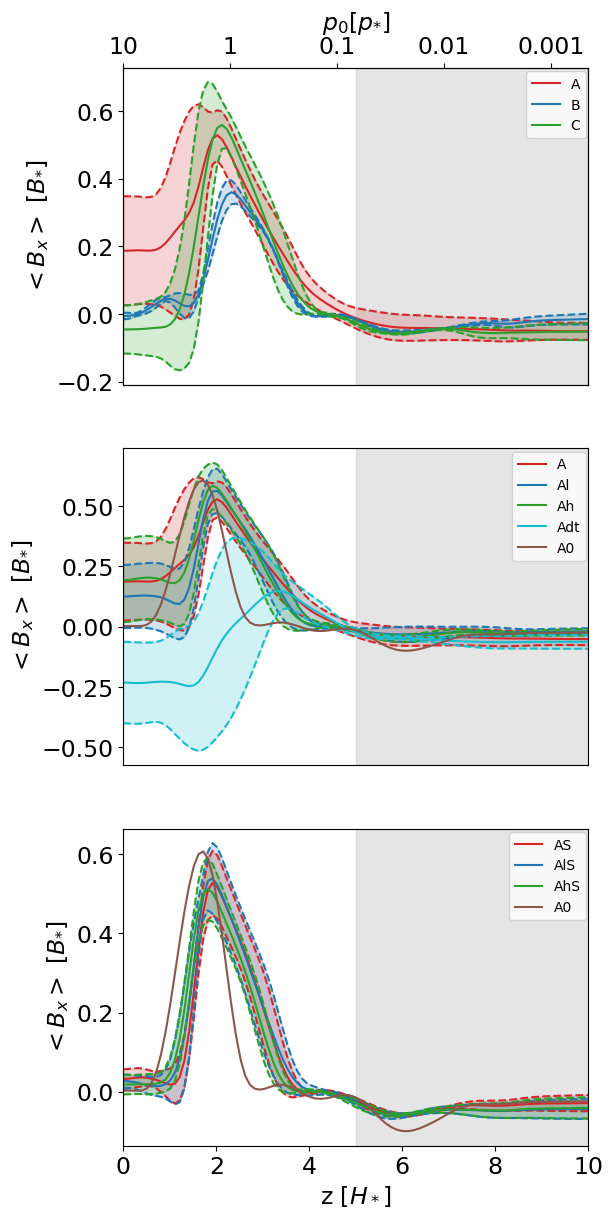}\\
\caption{ Comparison of the time-averaged vertical profiles $\langle\hat B_{x}\rangle({\hat z})$ averaged over the ${\hat x}$-${\hat y}$ plane, for different representative times. {\it Top panel:} simulations A, B and C. {\it Central panel:} simulations A0, A, Al, Ah and Adt. {\it Bottom panel}: A0, AS, AlS and AhS, which are all models without the local average of $\hat v_x$ in the forcing.  The solid lines indicate the average, and the dashed, with the enclosed shades, the standard deviation (except for the 1D model A0 for which the solution does not present stochastic variations).}.
\label{fig:results_wind_profiles}
\end{figure}

Let us now move to assess the dependence of the field configuration on the different wind profiles. The top panel of Fig. \ref{fig:results_wind_profiles} compares $\langle{\hat B}_{x}\rangle(z)$ for simulations A, B, and C. Firstly, the maximum value for $\langle{\hat B}_{x}\rangle(z)$ for simulations A and C is relatively similar, while for simulation B, the peak is slightly smaller and displaced to the right due to the increase of $z_{b}$ from 1.6 to 2. Furthermore, the deviations around the averaged profiles observed for simulations B and C are smaller compared to those in simulation A. Considering the standard deviations, the statistically significant differences are seen at the deepest side. Part of these deviations maybe to the different length of the simulations, with A lasting longer and experiencing a slow gradual rising of $\langle \hat B_x\rangle$ for $\hat z \lesssim 1$.

\subsection{Dependence on the forcing}

In the central panel of Fig.~\ref{fig:results_wind_profiles}, we compare the magnetic profiles (as above), for models A0 (no perturbations), A, Al, Ah, and Adt, i.e. varying either $\lambda$ (for the first three), or the timestep (At, which physically means a different time correlation between the random perturbations). The amplitude of the perturbation is not producing any substantial change in the average amplitude, meaning that the average state is not affected by the details of the turbulent motions. However, for the model Adt there is a much larger dispersion of values, with lower induced fields and currents.

In the bottom panel of Fig.~\ref{fig:results_wind_profiles} we show the same comparison for different $\lambda$, but in the case of the version of the forcing without the local averaging of $\hat v_x$, as discussed above. Such models have very similar average profiles of the models A0, A, Al and Ah, respectively. The differences among different perturbation amplitudes are negligible also in this case, as shown by the overlapping shaded regions. However, they present a much smaller dispersion of the values, reducing greatly the temporal changes, and giving a more constrained average profile as well. Considering also the model Adt, our interpretation is that rising the temporal correlation (lower $dt$) or spatial correlation (doing local average $\langle v_x \rangle$ in the forcing) of the induced perturbations makes the timescales of saturation longer, allowing stronger long-term pseudo-oscillatory changes in the configuration.

A comparison between A0 and the full 3D models shows that the presence of turbulence ($\lambda>0$) slightly shifts the dominant component $B_x$ in pressure (and the main component of the current, $J_y$), but the overall quantity (the peak of $B_x$) is of the same order. More importantly, turbulence adds a weaker, but non-zero, radial current component (which is otherwise zero), which might be relevant for the interaction with the deeper levels.

Summarizing, the details of the forcing employed can quantitatively (but not qualitatively) affect only to a minor extent the average profiles. Since the forcing is an effective way to mimic the effects of wind and turbulence, we can consider these model-to-model differences as fundamental model uncertainties, related to the nature of the perturbations.

\section{Applicability and estimate of the Ohmic dissipation}\label{sec:applicability}

The local simulations we have performed here are scale-invariant and we employ a fixed (local) profile of the wind, over which temperature and magnetic fields have no feedback (in other words, we don't have the full global pattern of thermal winds like in GCMs). Instead, in order to convert to physical units, we assign a-posteriori the reference values of temperature $T_0$, pressure $p_*$, gravity $g$ and average molecular weight $\mu$ (entering in the unit conversion factors, \S\ref{sec:rescaled_eqs}, via their dimensionless values $T_{2000}$, $p_{bar}$, $g_{10}$, $\mu$, respectively). While the latter two only modify the final physical units and can be assigned with the reasonable expected range, the pressure and temperature have to be chosen consistently with the choice of the wind profile and the isothermal assumption.

The reference pressure (i.e., the units of pressure $p_*$, which set its value at the bottom of our domain, $p_b$) is set to be consistent with the shear layer being located around $\sim 1$ bar, according to most GCM studies (although they usually assume deeper RCBs than what HJs should have, \citealt{thorngren19}). On the other hand, we know from GCMs that the vertical wind profile depends: (i) on the position (i.e.\ latitude and longitude), reaching the value of the speed of sound in the most irradiated (sub-stellar) outermost layers ($\lesssim 0.1$ bar); and (ii) on the temperature, because winds are powered by the thermal contrast via irradiation, and because magnetic dragging induced by strong thermal gradients can slow them down 
\citep{perna10a}. The resulting heating efficiency $\epsilon$, defined as the ratio of Ohmic heating to irradiation flux, peaks around $\sim 1500-1600$~K (e.g., \citealt{thorngren18}).

Although we don't consider the wind-temperature relations in this work, we can put some physical constraints on the range of temperatures $T_0$ that we can consider without violating physical constraints. First of all, ideal MHD can be applied if ${\cal R}m\gg 1$. The higher the temperature, the greater the ionization, and therefore, the higher the conductivity. This is granted for $T\gtrsim 2000$ K, according to \cite{dietrich22}.

Secondly, here we provide an additional self-consistency check for our local simulations. Seen from a macroscopic perspective, the amount of Ohmic dissipation should be much smaller than the irradiation flux. Previous works have considered a maximum global (i.e.\ integrated over the whole surface of the planet) conversion efficiency (ratio of Ohmic heating to irradiation) of about $\epsilon \lesssim 5\%$ \citep{batygin11}, since larger values of $\epsilon$ might make the planet exponentially unstable via atmospheric losses. \cite{thorngren18} used Bayesian statistics to infer the distribution of $\epsilon$ as a function of the incoming flux, obtaining typical values of $\lesssim 2-3 \%$ at most (dropping fast for higher $T$).

We do not have a global simulation, but we can compare the two local fluxes (energy per surface per time) as
\begin{equation}
    \int_z Q_j(z) dz < F_{irr} = \sigma_{sb}T_0^4 \simeq 0.9~T_{2000}^4 \frac{{\rm MW}}{{\rm m^{2}}}~, \label{eq:int_QJ}
\end{equation}
where the left-hand side is the heat flux released by Ohmic dissipation along the column, and $F_{irr}$ is the irradiation over the column under consideration ($\sigma_{sb}$ is the Stefan-Boltzmann constant, and the usual factor 4 in $F_{irr}$ is not present since this is not the globally averaged irradiation).

The electrical conductivity can be approximated as done e.g. in \cite{perna10a}, after a re-grouping of pre-coefficients and using the dimensionless value $T_{2000}$:
\begin{eqnarray}
\sigma(T,p) & = & \frac{x_e(T,p)}{\langle\sigma v\rangle_e (T)} \frac{e^2}{m_e} \label{eq:sigma} \\
x_e(T,p) & = & 7.7 \times 10^{-4} \left(\frac{a_{K}}{10^{-7}} \right)^{1/2} T_{2000}^{3/4}\nonumber\\
&& \times \left(\frac{n_{10}}{n_n(T,p)}\right)^{1/2} e^{-\alpha/T_{2000}} \\
\langle\sigma v\rangle_e (T) & = & 10^{-19} \left(\frac{128 k_{\rm B} T}{9 \pi m_e}\right)^{1/2} {\rm m}^2 = \nonumber\\
& = & 3.7\times 10^{-14}~T_{2000}^{1/2} ~{\rm m}^3 {\rm s}^{-1}
\end{eqnarray}
where: $m_e$ is the electron mass; $e$ the elementary charge; $n_n = \rho/m$ is the neutral number density, $n_{10}=3.62\times 10^{25}$ m$^{-3}$ is its value at $p=10$ bar and $T=2000$ K; $\alpha = 12.594$ is a numerical factor. The function $\langle\sigma v\rangle_e$ is the momentum transfer rate coefficient between electrons and neutrals (the dominant channel) taken from \cite{draine83} and references within (see their \S~III). The electron fraction is approximated by the ionization fraction of potassium as in eq.~(30) of \cite{balbus00}, which is valid as long as $x_e$ itself is less than the element abundance, $a_K$ ($\sim 10^{-7}$ in the Sun). This condition is approximately valid for $T\lesssim 2500$ K. At such high temperatures, moreover, other elements (Na and Ca in particular) contribute to the ionization and the approximation above overestimates the conductivity by a factor of a few, compared to the more complete calculations of e.g. \cite{kumar21} and \cite{dietrich22}. A more detailed assessment should actually solve the full Saha equation for a given composition (as done, e.g., in \citealt{rogers14b}). However, we consider that for a rough a-posteriori assessment, this approximation is good enough.

\begin{figure}
\centering \includegraphics[width=\linewidth]{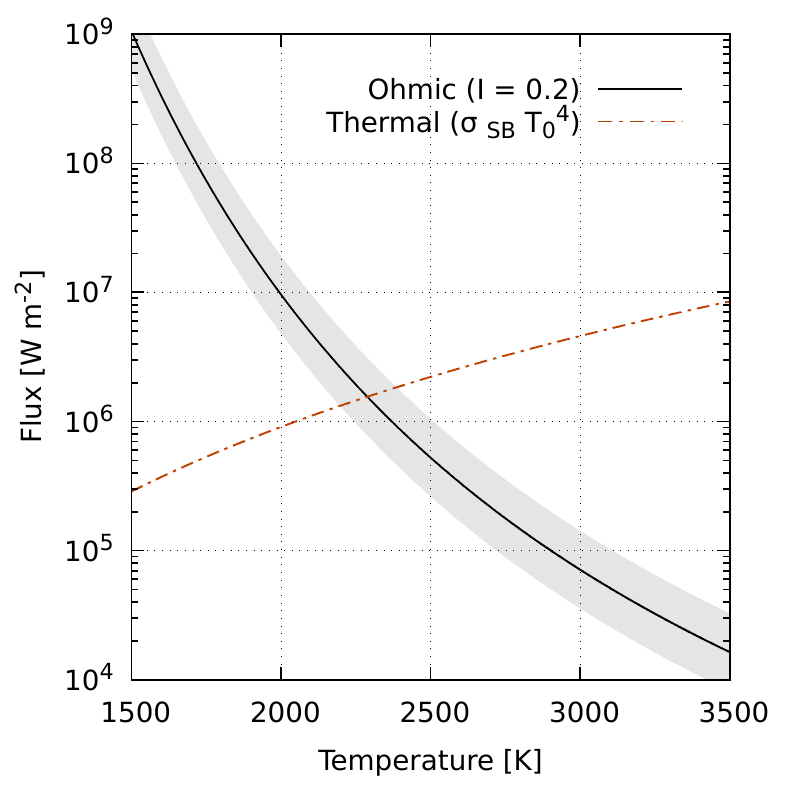}
\caption{Applicability of our results: conversion to physical units as a function of $T_0$, the background temperature that we assign a-posteriori to a given simulation. We compare the estimated Ohmic heat over the column (black line, assuming $g_{10}=1, \mu=2$, $p_{bar}=1$ and $a_k=10^{-7}$) and the local irradiation flux (red dot-dashed line). The gray shaded area indicates a range of typical values of the dimensionless integral ${\cal I}$, evaluated from the numerical simulations: from 0.1 to 0.4, while the black line is for 0.2. Our estimate is self-consistent only if the Ohmic heating is approximately below the irradiation line. From the plot, this corresponds to $T_0 \gtrsim 2300$~K, for the parameters chosen. Note that this plot is not meant to display the dependence of Ohmic dissipation on temperature: here, with 
our local simulations, we
consider a fixed wind profile and hence we do not account for the important global temperature-dependence on the wind (considered instead in the global simulations, GCMs).}
\label{fig:Ohmic_flux}
\end{figure}

Keeping this in mind, we can quantitatively estimate the local Ohmic dissipation a-posteriori, evaluating the physical conductivity and assuming that the results would not drastically change. Assuming an homogeneous composition in the domain considered, the only quantity of $\sigma(T)$ that depends on the height $\hat{z}$ (via pressure) is:
\begin{equation}
n_n(T,p(\hat{z})) = \frac{\rho(T,\hat{z})}{\mu m_u} \simeq \frac{p_b}{k_{\rm B} T_0} e^{-\hat{z}}~, \label{eq:n_n}
\end{equation}
where we have assumed the background state, i.e. $T\simeq T_0$ and $p_1 \ll p_0$. Then the $\hat{z}$-dependence in conductivity can be factorized by defining $\sigma(T)=\sigma_*(T)~e^{\hat{z}/2}$ (see App. \ref{app:diffusivity} for plots), such that the a-posteriori estimated Ohmic dissipation is:

\begin{equation}
Q_j(T) = \zeta(T)~{\cal I}~,
\label{eq:J2e}
\end{equation}
where we have defined a function that combines all of the temperature-dependent terms and numerical factors in eqs. (\ref{eq:sigma})-(\ref{eq:n_n}):
\begin{equation}
    \zeta(T)=\frac{J_*^2 H_*}{\sigma_*(T)} \simeq 24~\left( \frac{e^{\frac{\alpha}{T_{2000}}}}{T_{2000}^{7/4}}~\frac{\mu~ g_{10}}{a_{K,7}^{1/2}}\right)~\frac{{\rm MW}}{{\rm m^{2}}}, 
\end{equation}
where $a_{K,7}:=a_K/10^{-7}$, we have used $p_b = 10$ bar, eq.~(\ref{eq:jstar}) for $J_*$ and eq.~(\ref{eq:def_hp}) for $H_*$. The dimensionless integral
\begin{equation}
{\cal I} := \int_0^L \langle \hat{J}^2 \rangle ({\hat z}) e^{-\hat{z}/2} d\hat{z}
\label{eq:int_I}
\end{equation}
is numerically computed by using the average vertical profiles $\langle\hat{J}^2\rangle(z)$ shown above. In our numerical simulations, we typically obtain ${\cal I} \sim 0.1-0.3$ (see last column of Table~\ref{tab:models}). As said above, such integral physically depends on the temperature, since it determines the wind profile, which here we fix for a given simulation. When we use higher resolutions, as commented above, we obtain higher values of $B_x$, $J_y$, and, as a consequence, ${\cal I}$. In the 1D problem (A0), for $N_z=50,100,200,400$, we obtain the values of ${\cal I}=0.11,0.36,0.85,1.13$, respectively. This is consistent with the fact that higher resolutions correspond to lower numerical resistivities, which are representative of hotter Jupiters. The chosen resolution here ($N_z=100$) corresponds to an implicit numerical resistivity comparable to a physical one for very high temperatures ($T_0\gtrsim 3000$ K) at the pressures of the shear layer where most currents live (see Appendix~\ref{app:diffusivity}). In this range, local energy balance gives local efficiencies $Q_j/F_{irr} \lesssim 0.01$ (Fig. ~\ref{fig:Ohmic_flux}), which are reasonable values, comparable to the usual estimates of overall heating efficiencies \citep{thorngren19}. In a follow-up work, we will implement realistic profiles of the magnetic diffusivity for different temperatures. Since in most HJs ($T_{eq}\sim 1500-2500$ K) the diffusivity will be higher than the implicit one implemented here, we expect lower values of $B_x$ and $J_y$.

Since Fig.~\ref{fig:Ohmic_flux} only takes into account the dependencies on temperature of the unit conversion and the conductivity $\sigma(T)$, and not the wind $w(T)$, the decrease of Ohmic heating with temperature is steeper than in reality. For low temperatures, the wind and induced magnetic fields (hence, ${\cal I}$) are much lower, hence the curve would bend down. At high temperatures, the intensity of the wind saturates due to magnetic drag, so the trend could be more realistic. In fact, \cite{thorngren18} infer a strong decrease of efficiency (well below $1\%$) for very HJs.

 Finally, note that here we are only considering the energetics that represent a column in the hotter part of the day-side atmosphere. Hence, the temperature here considered is higher than the the equilibrium temperature (which is the one entering in the global energy balance), $T_0 > T_{eq}$. This also means that the shear-induced currents are on average lower than what we infer here, so the total Ohmic deposition will be less. As a matter of fact, the velocity profile changes according to the longitude, and the ideal MHD including winding mechanism and turbulence apply only in the regions with high enough ${\cal R}m$, i.e., in the regions with the steepest shear layers (e.g., \citealt{rogers14a}, \citealt{beltz22}).  Therefore, Fig.~\ref{fig:Ohmic_flux} should not be taken as an indication of the global energetics (efficiency), although it provides important qualitative insights of what happens at high $T$ (non-linear regimes).

\section{Final remarks}\label{sec:conclusions}

In this study we have performed ideal MHD simulations of a Hot Jupiter's narrow atmospheric column. We have used a forcing to mimic a realistic model of zonal wind profile, on which we added random perturbations. The common behaviour observed in the simulations is that the shear layer gives rise to an intense magnetic field winding, 
consistently with a semi-analytically estimate.

At equilibrium, such a toroidal field can reach local values of up to a few kG, in a very small region. This huge magnetic field is sustained by meridional currents that close within a narrow vertical layer, basically corresponding to the shear region. These results are in qualitative agreement with the non-linear regime by \cite{dietrich22} and share common features with the dynamo in Jovian outer layers simulated by \cite{wicht19a,wicht19b}, scaled by orders of magnitude due to the higher energy budget. 

The turbulence induced by the random perturbations creates an additional magnetic field, which tends to acquire a small meridional component (of the order of tens of Gauss maximum). The stratification hampers the stretching of the lines in the vertical (i.e. radial) direction, and the turbulent field tends to be evident in the bottom of the box, possibly pointing to a penetration of currents to deeper levels, a fundamental premise to power the radii inflation.

The results obtained show an important stochastic variability in the plane-averaged vertical profiles of the different quantities. A quantitative comparable variability comes from one model to another if the details of the effective forcing imposed is changed. However, varying the perturbation amplitude does not have strong effects. The position and width of the shear layer directly affect the shape of the induced currents, as expected. In any case, our main conclusions do not depend on these details, as long as the shear layer is steep enough.

Note that our results, interpreted here in terms of a zonal wind and induced toroidal field close to the substellar point, can apply also to a meridional wind: in that case the induced magnetic field would be mainly meridional, supported by local azimuthal currents.

In our ideal, scale-invariant MHD equations, we have an implicit constraint on the temperature that we assign a-posteriori to convert all values into physical units. As a matter of fact, we have neglected the resistivity, which is justifiable for sufficiently high values of the conductivity, $\sigma(T)$. As $\sigma$ increases with temperature, considerations of local energy balance, numerical dissipations and magnetic Reynlods number gives us a bound on the minimum local temperature of about $T_0 \gtrsim 3000$ K, above which these simulations can be considered applicable. Note that $T_0$ is higher than the global equilibrium temperature $T_{eq}$: if the redistribution is inefficient, (i.e., $T_0 \sim 2 T_{eq}$), then our results could be applied to the dayside columns of a good fraction of HJs ($T_{eq}\gtrsim 1800$ K).

The rough estimate of the Ohmic dissipation (via the a-posteriori calculation of the conductivity for a given $T$) can be used as a proxy for the deposited heat in the upper radiative layer. The inflation of radii is effective when the heat is deposited directly in the adiabat, below the RCB, which can be considerably shallow (at few bars) for very hot Jupiters \citep{thorngren19}. Considering these cases more carefully requires a re-definition of the background state, which we assumed as isothermal (i.e., well above the RCB) here. Moreover, GCMs (which inspire our choice of wind profiles) can give quite different results when the RCB is varied \citep{carone20}. Although we cannot infer the global energetics from our local simulations, we find that the  Ohmic heating efficiency (dissipated Ohmic energy vs. irradiation) decreases for increasing temperature, since the conductivity increases. Although here we do not consider the important dependence of wind on temperature (wind profiles are kept constant when we convert to physical units for a given $T$), such trends are probably real and compatible with other complementary studies \citep{thorngren18}, since the wind velocity stops increasing with $T$ due to magnetic drag \citep{perna10a}. Therefore, the results reported here should be regarded as an upper limit on the amount of Ohmic dissipation in the outer parts of the atmosphere (but not far from reality for very large $T$).

Besides these caveats, one of the most important results is the confinement of strong $\sim$ kG fields (likely less when real resistivity is considered) in the thin shear layer. Such confinement implies that they cannot unfortunately power detectable $\lesssim$ GHz coherent magnetospheric radio emission, expected via electron cyclotron maser mechanism at frequencies $\nu$ [MHz]=2.8 $B$ [G] \citep{zarka98}. However, the interaction between such shallow magnetic field and the one generated in the convective dynamo region, possibly reaching $\lesssim 100$ G \citep{cauley19}, is an interesting point to be explored further.

Our future aim is to extend the study to any temperature (i.e., applicable to any HJ), by including non-ideal effects, in particular the vertical profile of the finite conductivity and, possibly, assessing the relevance of Hall and ambipolar terms in presence of very large magnetic fields. Finally, we have used here a minimum resolution as a first study, but future simulations will employ high-resolution and a more detailed exploration of the forcing mechanism.
\section*{Acknowledgements}
CS's work has been carried out within the framework of the doctoral program in Physics of the Universitat Autònoma de Barcelona. CS, DV and TA are supported by the European Research Council (ERC) under the European Union’s Horizon 2020 research and innovation programme (ERC Starting Grant "IMAGINE" No. 948582, PI: DV). CS, DV, TA acknowledge support from ``Mar\'ia de Maeztu'' award to the Institut de Ciències de l'Espai (CEX2020-001058-M). We thank Fabio del Sordo and Albert Elias López for the useful comments and discussions, and Borja Miñano for the computational support. 

\section*{Data availability}
All data produced in this work will be shared on reasonable request
to the corresponding author.

\bibliographystyle{mnras}
\bibliography{references}

\begin{thebibliography}{}
\makeatletter
\relax
\def\mn@urlcharsother{\let\do\@makeother \do\$\do\&\do\#\do\^\do\_\do\%\do\~}
\def\mn@doi{\begingroup\mn@urlcharsother \@ifnextchar [ {\mn@doi@}
  {\mn@doi@[]}}
\def\mn@doi@[#1]#2{\def\@tempa{#1}\ifx\@tempa\@empty \href
  {http://dx.doi.org/#2} {doi:#2}\else \href {http://dx.doi.org/#2} {#1}\fi
  \endgroup}
\def\mn@eprint#1#2{\mn@eprint@#1:#2::\@nil}
\def\mn@eprint@arXiv#1{\href {http://arxiv.org/abs/#1} {{\tt arXiv:#1}}}
\def\mn@eprint@dblp#1{\href {http://dblp.uni-trier.de/rec/bibtex/#1.xml}
  {dblp:#1}}
\def\mn@eprint@#1:#2:#3:#4\@nil{\def\@tempa {#1}\def\@tempb {#2}\def\@tempc
  {#3}\ifx \@tempc \@empty \let \@tempc \@tempb \let \@tempb \@tempa \fi \ifx
  \@tempb \@empty \def\@tempb {arXiv}\fi \@ifundefined
  {mn@eprint@\@tempb}{\@tempb:\@tempc}{\expandafter \expandafter \csname
  mn@eprint@\@tempb\endcsname \expandafter{\@tempc}}}

\bibitem[\protect\citeauthoryear{{Arbona}, {Artigues}, {Bona-Casas},
  {Mass{\'o}}, {Mi{\~n}ano}, {Rigo}, {Trias}  \& {Bona}}{{Arbona}
  et~al.}{2013}]{2013CoPhC.184.2321A}
{Arbona} A.,  {Artigues} A.,  {Bona-Casas} C.,  {Mass{\'o}} J.,  {Mi{\~n}ano}
  B.,  {Rigo} A.,  {Trias} M.,   {Bona} C.,  2013, \mn@doi [Computer Physics
  Communications] {10.1016/j.cpc.2013.04.012}, \href
  {https://ui.adsabs.harvard.edu/abs/2013CoPhC.184.2321A} {184, 2321}

\bibitem[\protect\citeauthoryear{{Arbona}, {Mi{\~n}ano}, {Rigo}, {Bona},
  {Palenzuela}, {Artigues}, {Bona-Casas}  \& {Mass{\'o}}}{{Arbona}
  et~al.}{2018}]{2018CoPhC.229..170A}
{Arbona} A.,  {Mi{\~n}ano} B.,  {Rigo} A.,  {Bona} C.,  {Palenzuela} C.,
  {Artigues} A.,  {Bona-Casas} C.,   {Mass{\'o}} J.,  2018, \mn@doi [Computer
  Physics Communications] {10.1016/j.cpc.2018.03.015}, \href
  {https://ui.adsabs.harvard.edu/abs/2018CoPhC.229..170A} {229, 170}

\bibitem[\protect\citeauthoryear{{Balbus} \& {Hawley}}{{Balbus} \&
  {Hawley}}{2000}]{balbus00}
{Balbus} S.~A.,  {Hawley} J.~F.,  2000, \mn@doi [\ssr]
  {10.1023/A:1005293132737}, \href
  {https://ui.adsabs.harvard.edu/abs/2000SSRv...92...39B} {92, 39}

\bibitem[\protect\citeauthoryear{{Batygin} \& {Stevenson}}{{Batygin} \&
  {Stevenson}}{2010}]{batygin10}
{Batygin} K.,  {Stevenson} D.~J.,  2010, \mn@doi [\apjl]
  {10.1088/2041-8205/714/2/L238}, \href
  {https://ui.adsabs.harvard.edu/abs/2010ApJ...714L.238B} {714, L238}

\bibitem[\protect\citeauthoryear{{Batygin}, {Stevenson}  \&
  {Bodenheimer}}{{Batygin} et~al.}{2011}]{batygin11}
{Batygin} K.,  {Stevenson} D.~J.,   {Bodenheimer} P.~H.,  2011, \mn@doi [\apj]
  {10.1088/0004-637X/738/1/1}, \href
  {https://ui.adsabs.harvard.edu/abs/2011ApJ...738....1B} {738, 1}

\bibitem[\protect\citeauthoryear{{Batygin}, {Stanley}  \&
  {Stevenson}}{{Batygin} et~al.}{2013}]{batygin13}
{Batygin} K.,  {Stanley} S.,   {Stevenson} D.~J.,  2013, \mn@doi [\apj]
  {10.1088/0004-637X/776/1/53}, \href
  {https://ui.adsabs.harvard.edu/abs/2013ApJ...776...53B} {776, 53}

\bibitem[\protect\citeauthoryear{{Beltz}, {Rauscher}, {Kempton}, {Malsky},
  {Ochs}, {Arora}  \& {Savel}}{{Beltz} et~al.}{2022}]{beltz22}
{Beltz} H.,  {Rauscher} E.,  {Kempton} E. M.~R.,  {Malsky} I.,  {Ochs} G.,
  {Arora} M.,   {Savel} A.,  2022, \mn@doi [\aj] {10.3847/1538-3881/ac897b},
  \href {https://ui.adsabs.harvard.edu/abs/2022AJ....164..140B} {164, 140}

\bibitem[\protect\citeauthoryear{{Bodenheimer}, {Lin}  \&
  {Mardling}}{{Bodenheimer} et~al.}{2001}]{bodenheimer01}
{Bodenheimer} P.,  {Lin} D.~N.~C.,   {Mardling} R.~A.,  2001, \mn@doi [\apj]
  {10.1086/318667}, \href
  {https://ui.adsabs.harvard.edu/abs/2001ApJ...548..466B} {548, 466}

\bibitem[\protect\citeauthoryear{{Burrows}, {Hubeny}, {Budaj}  \&
  {Hubbard}}{{Burrows} et~al.}{2007}]{burrows07}
{Burrows} A.,  {Hubeny} I.,  {Budaj} J.,   {Hubbard} W.~B.,  2007, \mn@doi
  [\apj] {10.1086/514326}, \href
  {https://ui.adsabs.harvard.edu/abs/2007ApJ...661..502B} {661, 502}

\bibitem[\protect\citeauthoryear{{Carone} et~al.,}{{Carone}
  et~al.}{2020}]{carone20}
{Carone} L.,  et~al., 2020, \mn@doi [\mnras] {10.1093/mnras/staa1733}, \href
  {https://ui.adsabs.harvard.edu/abs/2020MNRAS.496.3582C} {496, 3582}

\bibitem[\protect\citeauthoryear{{Cauley}, {Shkolnik}, {Llama}  \&
  {Lanza}}{{Cauley} et~al.}{2019}]{cauley19}
{Cauley} P.~W.,  {Shkolnik} E.~L.,  {Llama} J.,   {Lanza} A.~F.,  2019, \mn@doi
  [Nature Astronomy] {10.1038/s41550-019-0840-x}, \href
  {https://ui.adsabs.harvard.edu/abs/2019NatAs...3.1128C} {3, 1128}

\bibitem[\protect\citeauthoryear{{Chabrier} \& {Baraffe}}{{Chabrier} \&
  {Baraffe}}{2007}]{chabrier07}
{Chabrier} G.,  {Baraffe} I.,  2007, \mn@doi [\apjl] {10.1086/518473}, \href
  {https://ui.adsabs.harvard.edu/abs/2007ApJ...661L..81C} {661, L81}

\bibitem[\protect\citeauthoryear{{Cho}, {Menou}, {Hansen}  \& {Seager}}{{Cho}
  et~al.}{2008}]{cho08}
{Cho} J. Y.~K.,  {Menou} K.,  {Hansen} B. M.~S.,   {Seager} S.,  2008, \mn@doi
  [\apj] {10.1086/524718}, \href
  {https://ui.adsabs.harvard.edu/abs/2008ApJ...675..817C} {675, 817}

\bibitem[\protect\citeauthoryear{{Dedner}, {Kemm}, {Kr{\"o}ner}, {Munz},
  {Schnitzer}  \& {Wesenberg}}{{Dedner} et~al.}{2002}]{dedner02}
{Dedner} A.,  {Kemm} F.,  {Kr{\"o}ner} D.,  {Munz} C.-D.,  {Schnitzer} T.,
  {Wesenberg} M.,  2002, \mn@doi [Journal of Computational Physics]
  {10.1006/jcph.2001.6961}, \href
  {http://adsabs.harvard.edu/abs/2002JCoPh.175..645D} {175, 645}

\bibitem[\protect\citeauthoryear{{Dietrich}, {Kumar}, {Poser}, {French},
  {Nettelmann}, {Redmer}  \& {Wicht}}{{Dietrich} et~al.}{2022}]{dietrich22}
{Dietrich} W.,  {Kumar} S.,  {Poser} A.~J.,  {French} M.,  {Nettelmann} N.,
  {Redmer} R.,   {Wicht} J.,  2022, \mn@doi [\mnras] {10.1093/mnras/stac2849},
  \href {https://ui.adsabs.harvard.edu/abs/2022MNRAS.517.3113D} {517, 3113}

\bibitem[\protect\citeauthoryear{{Dobbs-Dixon} \& {Lin}}{{Dobbs-Dixon} \&
  {Lin}}{2008}]{dobbs08}
{Dobbs-Dixon} I.,  {Lin} D.~N.~C.,  2008, \mn@doi [\apj] {10.1086/523786},
  \href {https://ui.adsabs.harvard.edu/abs/2008ApJ...673..513D} {673, 513}

\bibitem[\protect\citeauthoryear{{Draine}, {Roberge}  \& {Dalgarno}}{{Draine}
  et~al.}{1983}]{draine83}
{Draine} B.~T.,  {Roberge} W.~G.,   {Dalgarno} A.,  1983, \mn@doi [\apj]
  {10.1086/160617}, \href
  {https://ui.adsabs.harvard.edu/abs/1983ApJ...264..485D} {264, 485}

\bibitem[\protect\citeauthoryear{{Felipe}}{{Felipe}}{2010}]{tobias_thesis}
{Felipe} T.,  2010, PhD thesis, University of La Laguna, Spain

\bibitem[\protect\citeauthoryear{{Fortney}, {Dawson}  \& {Komacek}}{{Fortney}
  et~al.}{2021}]{fortney21}
{Fortney} J.~J.,  {Dawson} R.~I.,   {Komacek} T.~D.,  2021, \mn@doi [Journal of
  Geophysical Research (Planets)] {10.1029/2020JE006629}, \href
  {https://ui.adsabs.harvard.edu/abs/2021JGRE..12606629F} {126, e06629}

\bibitem[\protect\citeauthoryear{{Guillot}}{{Guillot}}{2010}]{guillot10}
{Guillot} T.,  2010, \mn@doi [\aap] {10.1051/0004-6361/200913396}, \href
  {https://ui.adsabs.harvard.edu/abs/2010A&A...520A..27G} {520, A27}

\bibitem[\protect\citeauthoryear{Gunney \& Anderson}{Gunney \&
  Anderson}{2016}]{gunney16}
Gunney B.~T.,  Anderson R.~W.,  2016, \mn@doi [Journal of Parallel and
  Distributed Computing] {https://doi.org/10.1016/j.jpdc.2015.11.005}, 89, 65

\bibitem[\protect\citeauthoryear{{Heng} \& {Showman}}{{Heng} \&
  {Showman}}{2015}]{heng15}
{Heng} K.,  {Showman} A.~P.,  2015, \mn@doi [Annual Review of Earth and
  Planetary Sciences] {10.1146/annurev-earth-060614-105146}, \href
  {https://ui.adsabs.harvard.edu/abs/2015AREPS..43..509H} {43, 509}

\bibitem[\protect\citeauthoryear{{Heng}, {Menou}  \& {Phillipps}}{{Heng}
  et~al.}{2011}]{heng11}
{Heng} K.,  {Menou} K.,   {Phillipps} P.~J.,  2011, \mn@doi [\mnras]
  {10.1111/j.1365-2966.2011.18315.x}, \href
  {https://ui.adsabs.harvard.edu/abs/2011MNRAS.413.2380H} {413, 2380}

\bibitem[\protect\citeauthoryear{Hornung \& Kohn}{Hornung \&
  Kohn}{2002}]{hornung02}
Hornung R.~D.,  Kohn S.~R.,  2002, \mn@doi [Concurrency and Computation:
  Practice and Experience] {10.1002/cpe.652}, 14, 347

\bibitem[\protect\citeauthoryear{{Kataria}, {Showman}, {Fortney}, {Stevenson},
  {Line}, {Kreidberg}, {Bean}  \& {D{\'e}sert}}{{Kataria}
  et~al.}{2015}]{kataria15}
{Kataria} T.,  {Showman} A.~P.,  {Fortney} J.~J.,  {Stevenson} K.~B.,  {Line}
  M.~R.,  {Kreidberg} L.,  {Bean} J.~L.,   {D{\'e}sert} J.-M.,  2015, \mn@doi
  [\apj] {10.1088/0004-637X/801/2/86}, \href
  {https://ui.adsabs.harvard.edu/abs/2015ApJ...801...86K} {801, 86}

\bibitem[\protect\citeauthoryear{{Knutson} et~al.,}{{Knutson}
  et~al.}{2007}]{knutson07}
{Knutson} H.~A.,  et~al., 2007, \mn@doi [\nat] {10.1038/nature05782}, \href
  {https://ui.adsabs.harvard.edu/abs/2007Natur.447..183K} {447, 183}

\bibitem[\protect\citeauthoryear{{Knutson} et~al.,}{{Knutson}
  et~al.}{2012}]{knutson12}
{Knutson} H.~A.,  et~al., 2012, \mn@doi [\apj] {10.1088/0004-637X/754/1/22},
  \href {https://ui.adsabs.harvard.edu/abs/2012ApJ...754...22K} {754, 22}

\bibitem[\protect\citeauthoryear{{Koll} \& {Komacek}}{{Koll} \&
  {Komacek}}{2018}]{koll18}
{Koll} D. D.~B.,  {Komacek} T.~D.,  2018, \mn@doi [\apj]
  {10.3847/1538-4357/aaa3de}, \href
  {https://ui.adsabs.harvard.edu/abs/2018ApJ...853..133K} {853, 133}

\bibitem[\protect\citeauthoryear{{Komacek} \& {Youdin}}{{Komacek} \&
  {Youdin}}{2017}]{komacek17}
{Komacek} T.~D.,  {Youdin} A.~N.,  2017, \mn@doi [\apj]
  {10.3847/1538-4357/aa7b75}, \href
  {https://ui.adsabs.harvard.edu/abs/2017ApJ...844...94K} {844, 94}

\bibitem[\protect\citeauthoryear{{Komacek}, {Thorngren}, {Lopez}  \&
  {Ginzburg}}{{Komacek} et~al.}{2020}]{komacek20}
{Komacek} T.~D.,  {Thorngren} D.~P.,  {Lopez} E.~D.,   {Ginzburg} S.,  2020,
  \mn@doi [\apj] {10.3847/1538-4357/ab7eb4}, \href
  {https://ui.adsabs.harvard.edu/abs/2020ApJ...893...36K} {893, 36}

\bibitem[\protect\citeauthoryear{{Komacek}, {Tan}, {Gao}  \& {Lee}}{{Komacek}
  et~al.}{2022}]{komacek22}
{Komacek} T.~D.,  {Tan} X.,  {Gao} P.,   {Lee} E. K.~H.,  2022, \mn@doi [\apj]
  {10.3847/1538-4357/ac7723}, \href
  {https://ui.adsabs.harvard.edu/abs/2022ApJ...934...79K} {934, 79}

\bibitem[\protect\citeauthoryear{{Kumar}, {Poser}, {Sch{\"o}ttler},
  {Kleinschmidt}, {Dietrich}, {Wicht}, {French}  \& {Redmer}}{{Kumar}
  et~al.}{2021}]{kumar21}
{Kumar} S.,  {Poser} A.~J.,  {Sch{\"o}ttler} M.,  {Kleinschmidt} U.,
  {Dietrich} W.,  {Wicht} J.,  {French} M.,   {Redmer} R.,  2021, \mn@doi
  [\pre] {10.1103/PhysRevE.103.063203}, \href
  {https://ui.adsabs.harvard.edu/abs/2021PhRvE.103f3203K} {103, 063203}

\bibitem[\protect\citeauthoryear{{Laughlin}, {Crismani}  \& {Adams}}{{Laughlin}
  et~al.}{2011}]{laughlin11}
{Laughlin} G.,  {Crismani} M.,   {Adams} F.~C.,  2011, \mn@doi [\apjl]
  {10.1088/2041-8205/729/1/L7}, \href
  {https://ui.adsabs.harvard.edu/abs/2011ApJ...729L...7L} {729, L7}

\bibitem[\protect\citeauthoryear{{Li} \& {Goodman}}{{Li} \&
  {Goodman}}{2010}]{li10}
{Li} J.,  {Goodman} J.,  2010, \mn@doi [\apj] {10.1088/0004-637X/725/1/1146},
  \href {https://ui.adsabs.harvard.edu/abs/2010ApJ...725.1146L} {725, 1146}

\bibitem[\protect\citeauthoryear{{Liu}, {Goldreich}  \& {Stevenson}}{{Liu}
  et~al.}{2008}]{liu08}
{Liu} J.,  {Goldreich} P.~M.,   {Stevenson} D.~J.,  2008, \mn@doi [\icarus]
  {10.1016/j.icarus.2007.11.036}, \href
  {https://ui.adsabs.harvard.edu/abs/2008Icar..196..653L} {196, 653}

\bibitem[\protect\citeauthoryear{{Lopez} \& {Fortney}}{{Lopez} \&
  {Fortney}}{2016}]{lopez16}
{Lopez} E.~D.,  {Fortney} J.~J.,  2016, \mn@doi [\apj]
  {10.3847/0004-637X/818/1/4}, \href
  {https://ui.adsabs.harvard.edu/abs/2016ApJ...818....4L} {818, 4}

\bibitem[\protect\citeauthoryear{{Menou}}{{Menou}}{2020}]{menou20}
{Menou} K.,  2020, \mn@doi [\mnras] {10.1093/mnras/staa532}, \href
  {https://ui.adsabs.harvard.edu/abs/2020MNRAS.493.5038M} {493, 5038}

\bibitem[\protect\citeauthoryear{{Menou}}{{Menou}}{2022}]{menou22}
{Menou} K.,  2022, \mn@doi [\mnras] {10.1093/mnras/stac2854}, \href
  {https://ui.adsabs.harvard.edu/abs/2022MNRAS.517.2714M} {517, 2714}

\bibitem[\protect\citeauthoryear{{Palenzuela} et~al.,}{{Palenzuela}
  et~al.}{2018}]{palenzuela18}
{Palenzuela} C.,  et~al., 2018, \mn@doi [Classical and Quantum Gravity]
  {10.1088/1361-6382/aad7f6}, \href
  {https://ui.adsabs.harvard.edu/abs/2018CQGra..35r5007P} {35, 185007}

\bibitem[\protect\citeauthoryear{{Palenzuela}, {Mi{\~n}ano}, {Arbona},
  {Bona-Casas}, {Bona}  \& {Mass{\'o}}}{{Palenzuela}
  et~al.}{2021}]{palenzuela21}
{Palenzuela} C.,  {Mi{\~n}ano} B.,  {Arbona} A.,  {Bona-Casas} C.,  {Bona} C.,
   {Mass{\'o}} J.,  2021, \mn@doi [Computer Physics Communications]
  {10.1016/j.cpc.2020.107675}, \href
  {https://ui.adsabs.harvard.edu/abs/2021CoPhC.25907675P} {259, 107675}

\bibitem[\protect\citeauthoryear{{Parmentier}, {Showman}  \&
  {Lian}}{{Parmentier} et~al.}{2013}]{parmentier13}
{Parmentier} V.,  {Showman} A.~P.,   {Lian} Y.,  2013, \mn@doi [\aap]
  {10.1051/0004-6361/201321132}, \href
  {https://ui.adsabs.harvard.edu/abs/2013A&A...558A..91P} {558, A91}

\bibitem[\protect\citeauthoryear{{Parmentier} et~al.,}{{Parmentier}
  et~al.}{2018}]{parmentier18}
{Parmentier} V.,  et~al., 2018, \mn@doi [\aap] {10.1051/0004-6361/201833059},
  \href {https://ui.adsabs.harvard.edu/abs/2018A&A...617A.110P} {617, A110}

\bibitem[\protect\citeauthoryear{{Perez-Becker} \& {Showman}}{{Perez-Becker} \&
  {Showman}}{2013}]{perez13}
{Perez-Becker} D.,  {Showman} A.~P.,  2013, \mn@doi [\apj]
  {10.1088/0004-637X/776/2/134}, \href
  {https://ui.adsabs.harvard.edu/abs/2013ApJ...776..134P} {776, 134}

\bibitem[\protect\citeauthoryear{{Perna}, {Menou}  \& {Rauscher}}{{Perna}
  et~al.}{2010a}]{perna10a}
{Perna} R.,  {Menou} K.,   {Rauscher} E.,  2010a, \mn@doi [\apj]
  {10.1088/0004-637X/719/2/1421}, \href
  {https://ui.adsabs.harvard.edu/abs/2010ApJ...719.1421P} {719, 1421}

\bibitem[\protect\citeauthoryear{{Perna}, {Menou}  \& {Rauscher}}{{Perna}
  et~al.}{2010b}]{perna10b}
{Perna} R.,  {Menou} K.,   {Rauscher} E.,  2010b, \mn@doi [\apj]
  {10.1088/0004-637X/724/1/313}, \href
  {https://ui.adsabs.harvard.edu/abs/2010ApJ...724..313P} {724, 313}

\bibitem[\protect\citeauthoryear{{Perna}, {Heng}  \& {Pont}}{{Perna}
  et~al.}{2012}]{perna12}
{Perna} R.,  {Heng} K.,   {Pont} F.,  2012, \mn@doi [\apj]
  {10.1088/0004-637X/751/1/59}, \href
  {https://ui.adsabs.harvard.edu/abs/2012ApJ...751...59P} {751, 59}

\bibitem[\protect\citeauthoryear{{Rauscher} \& {Menou}}{{Rauscher} \&
  {Menou}}{2012}]{rauscher12}
{Rauscher} E.,  {Menou} K.,  2012, \mn@doi [\apj] {10.1088/0004-637X/750/2/96},
  \href {https://ui.adsabs.harvard.edu/abs/2012ApJ...750...96R} {750, 96}

\bibitem[\protect\citeauthoryear{{Rauscher} \& {Menou}}{{Rauscher} \&
  {Menou}}{2013}]{rauscher13}
{Rauscher} E.,  {Menou} K.,  2013, \mn@doi [\apj]
  {10.1088/0004-637X/764/1/103}, \href
  {https://ui.adsabs.harvard.edu/abs/2013ApJ...764..103R} {764, 103}

\bibitem[\protect\citeauthoryear{{Rogers} \& {Komacek}}{{Rogers} \&
  {Komacek}}{2014}]{rogers14b}
{Rogers} T.~M.,  {Komacek} T.~D.,  2014, \mn@doi [\apj]
  {10.1088/0004-637X/794/2/132}, \href
  {https://ui.adsabs.harvard.edu/abs/2014ApJ...794..132R} {794, 132}

\bibitem[\protect\citeauthoryear{{Rogers} \& {McElwaine}}{{Rogers} \&
  {McElwaine}}{2017}]{rogers17}
{Rogers} T.~M.,  {McElwaine} J.~N.,  2017, \mn@doi [\apjl]
  {10.3847/2041-8213/aa72da}, \href
  {https://ui.adsabs.harvard.edu/abs/2017ApJ...841L..26R} {841, L26}

\bibitem[\protect\citeauthoryear{{Rogers} \& {Showman}}{{Rogers} \&
  {Showman}}{2014}]{rogers14a}
{Rogers} T.~M.,  {Showman} A.~P.,  2014, \mn@doi [\apjl]
  {10.1088/2041-8205/782/1/L4}, \href
  {https://ui.adsabs.harvard.edu/abs/2014ApJ...782L...4R} {782, L4}

\bibitem[\protect\citeauthoryear{{Ryu}, {Zingales}  \& {Perna}}{{Ryu}
  et~al.}{2018}]{ryu18}
{Ryu} T.,  {Zingales} M.,   {Perna} R.,  2018, \mn@doi [\mnras]
  {10.1093/mnras/sty2638}, \href
  {https://ui.adsabs.harvard.edu/abs/2018MNRAS.481.5517R} {481, 5517}

\bibitem[\protect\citeauthoryear{{Showman} \& {Guillot}}{{Showman} \&
  {Guillot}}{2002}]{showman02}
{Showman} A.~P.,  {Guillot} T.,  2002, \mn@doi [\aap]
  {10.1051/0004-6361:20020101}, \href
  {https://ui.adsabs.harvard.edu/abs/2002A&A...385..166S} {385, 166}

\bibitem[\protect\citeauthoryear{{Showman}, {Fortney}, {Lian}, {Marley},
  {Freedman}, {Knutson}  \& {Charbonneau}}{{Showman} et~al.}{2009}]{showman09}
{Showman} A.~P.,  {Fortney} J.~J.,  {Lian} Y.,  {Marley} M.~S.,  {Freedman}
  R.~S.,  {Knutson} H.~A.,   {Charbonneau} D.,  2009, \mn@doi [\apj]
  {10.1088/0004-637X/699/1/564}, \href
  {https://ui.adsabs.harvard.edu/abs/2009ApJ...699..564S} {699, 564}

\bibitem[\protect\citeauthoryear{{Showman}, {Cho}  \& {Menou}}{{Showman}
  et~al.}{2010}]{showman10}
{Showman} A.~P.,  {Cho} J.~Y.~K.,   {Menou} K.,  2010, in {Seager} S.,  ed., ,
  Exoplanets.
pp 471--516, \mn@doi{10.48550/arXiv.0911.3170}

\bibitem[\protect\citeauthoryear{{Showman}, {Lewis}  \& {Fortney}}{{Showman}
  et~al.}{2015}]{showman15}
{Showman} A.~P.,  {Lewis} N.~K.,   {Fortney} J.~J.,  2015, \mn@doi [\apj]
  {10.1088/0004-637X/801/2/95}, \href
  {https://ui.adsabs.harvard.edu/abs/2015ApJ...801...95S} {801, 95}

\bibitem[\protect\citeauthoryear{{Snellen}, {de Kok}, {de Mooij}  \&
  {Albrecht}}{{Snellen} et~al.}{2010}]{snellen10}
{Snellen} I. A.~G.,  {de Kok} R.~J.,  {de Mooij} E. J.~W.,   {Albrecht} S.,
  2010, \mn@doi [\nat] {10.1038/nature09111}, \href
  {https://ui.adsabs.harvard.edu/abs/2010Natur.465.1049S} {465, 1049}

\bibitem[\protect\citeauthoryear{{Thorngren} \& {Fortney}}{{Thorngren} \&
  {Fortney}}{2018}]{thorngren18}
{Thorngren} D.~P.,  {Fortney} J.~J.,  2018, \mn@doi [\aj]
  {10.3847/1538-3881/aaba13}, \href
  {https://ui.adsabs.harvard.edu/abs/2018AJ....155..214T} {155, 214}

\bibitem[\protect\citeauthoryear{{Thorngren}, {Gao}  \& {Fortney}}{{Thorngren}
  et~al.}{2019}]{thorngren19}
{Thorngren} D.,  {Gao} P.,   {Fortney} J.~J.,  2019, \mn@doi [\apjl]
  {10.3847/2041-8213/ab43d0}, \href
  {https://ui.adsabs.harvard.edu/abs/2019ApJ...884L...6T} {884, L6}

\bibitem[\protect\citeauthoryear{{Vigan{\`o}}, {Aguilera-Miret}  \&
  {Palenzuela}}{{Vigan{\`o}} et~al.}{2019}]{vigano19}
{Vigan{\`o}} D.,  {Aguilera-Miret} R.,   {Palenzuela} C.,  2019, \mn@doi
  [Physics of Fluids] {10.1063/1.5121546}, \href
  {https://ui.adsabs.harvard.edu/abs/2019PhFl...31j5102V} {31, 105102}

\bibitem[\protect\citeauthoryear{{Wang}, {Fischer}, {Horch}  \& {Huang}}{{Wang}
  et~al.}{2015}]{wang15}
{Wang} J.,  {Fischer} D.~A.,  {Horch} E.~P.,   {Huang} X.,  2015, \mn@doi
  [\apj] {10.1088/0004-637X/799/2/229}, \href
  {https://ui.adsabs.harvard.edu/abs/2015ApJ...799..229W} {799, 229}

\bibitem[\protect\citeauthoryear{{Wicht}, {Gastine}  \& {Duarte}}{{Wicht}
  et~al.}{2019a}]{wicht19a}
{Wicht} J.,  {Gastine} T.,   {Duarte} L.~D.~V.,  2019a, \mn@doi [Journal of
  Geophysical Research (Planets)] {10.1029/2018JE005759}, \href
  {https://ui.adsabs.harvard.edu/abs/2019JGRE..124..837W} {124, 837}

\bibitem[\protect\citeauthoryear{{Wicht}, {Gastine}, {Duarte}  \&
  {Dietrich}}{{Wicht} et~al.}{2019b}]{wicht19b}
{Wicht} J.,  {Gastine} T.,  {Duarte} L.~D.~V.,   {Dietrich} W.,  2019b, \mn@doi
  [\aap] {10.1051/0004-6361/201935682}, \href
  {https://ui.adsabs.harvard.edu/abs/2019A&A...629A.125W} {629, A125}

\bibitem[\protect\citeauthoryear{{Youdin} \& {Mitchell}}{{Youdin} \&
  {Mitchell}}{2010}]{youdin10}
{Youdin} A.~N.,  {Mitchell} J.~L.,  2010, \mn@doi [\apj]
  {10.1088/0004-637X/721/2/1113}, \href
  {https://ui.adsabs.harvard.edu/abs/2010ApJ...721.1113Y} {721, 1113}

\bibitem[\protect\citeauthoryear{{Zarka}}{{Zarka}}{1998}]{zarka98}
{Zarka} P.,  1998, \mn@doi [\jgr] {10.1029/98JE01323}, \href
  {https://ui.adsabs.harvard.edu/abs/1998JGR...10320159Z} {103, 20159}

\bibitem[\protect\citeauthoryear{{Zingale} et~al.,}{{Zingale}
  et~al.}{2002}]{zingale02}
{Zingale} M.,  et~al., 2002, \mn@doi [\apjs] {10.1086/342754}, \href
  {https://ui.adsabs.harvard.edu/abs/2002ApJS..143..539Z} {143, 539}

\makeatother
\end{thebibliography}
\appendix
\section{Hydrostatic stability}\label{app:stability_tests}

We have tested the hydrostatic stability of the atmospheric column represented by our domain. The problem is typical of different astrophysical scenarios, where one needs to simulate fast dynamics over a much longer timescale. Specific solutions have been studied (e.g., \citealt{zingale02}), depending on the numerical scheme employed. In our case, the total timescale is the one needed to reach a stationary state, which under our conditions is typically hundreds time the sound crossing time.

The basic test is then to study the numerical stability of the vertical background profile described in \S\ref{sec:domain}. We run simulations with no forcing and no initial velocity perturbations or magnetic fields, $\vec{v}(t=0)=\vec{B}(t=0)=\vec{F}_{\rm src}=0$. Such tests are 1D, since the only spatial dependence in the initial fields is on $z$. We then analyse the results computing the domain-integrated L2-norm of $\hat v_z$, i.e. the ambient vertical velocities that depart from the analytical equilibrium ($\hat v_z=\hat\rho_1=\hat p_1=0$).

In all cases, tiny vertical velocities arise from the boundaries and from the readjustment due to discretisation. Part of it is due to the huge differences in the magnitudes of the fields, an issue that is greatly mitigated by evolving the perturbed density $\rho_1$ instead of the total one $\rho$. In order to find the numerical setup that minimizes such spurious velocities, we explore the influence of the following numerical parameters: (i) $\nu_{\rm cool}$; (ii) vertical size of the domain ($L$); (iii) parameters of the artificial damping region: lower boundary ($z_d$) and amplitude ($A_d$); (iv) resolution.

\begin{figure}
\centering
\includegraphics[trim={0 1.7cm 3cm 1.65cm},clip,width=\linewidth]{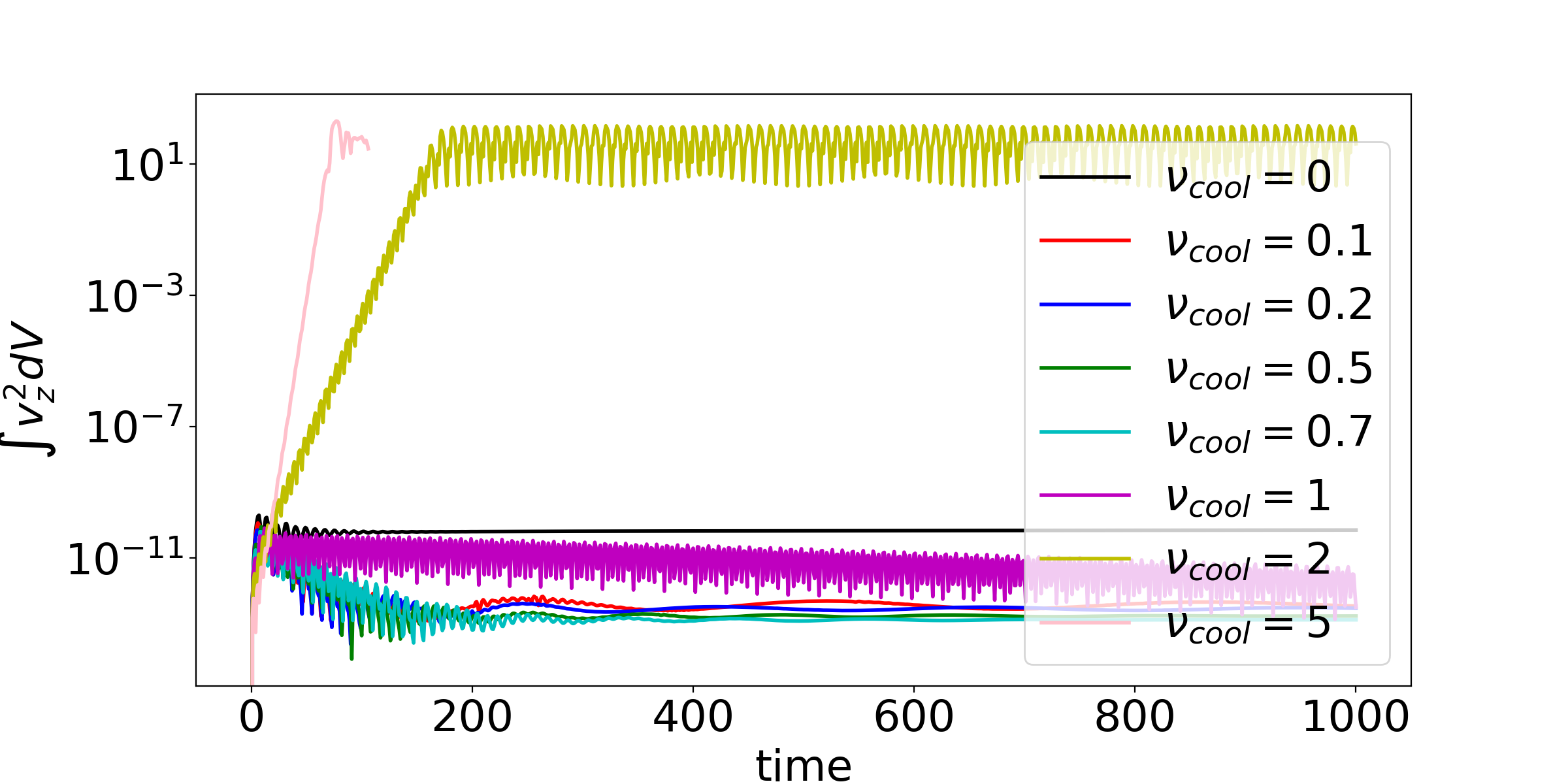}\\
\includegraphics[trim={0 1.7cm 3cm 1.65cm},clip,width=\linewidth]{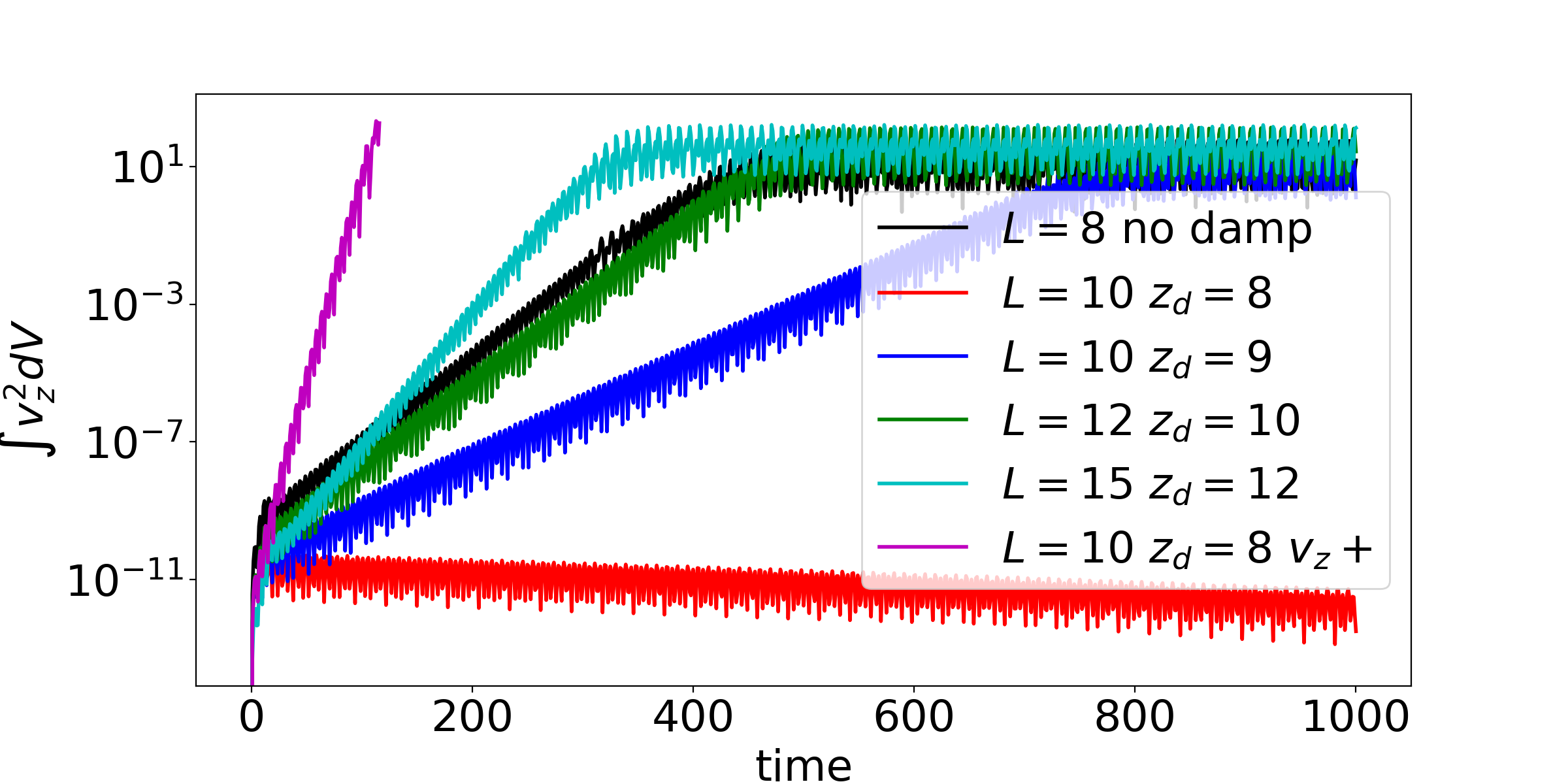}\\
\includegraphics[trim={0 1.7cm 3cm 1.65cm},clip,width=\linewidth]{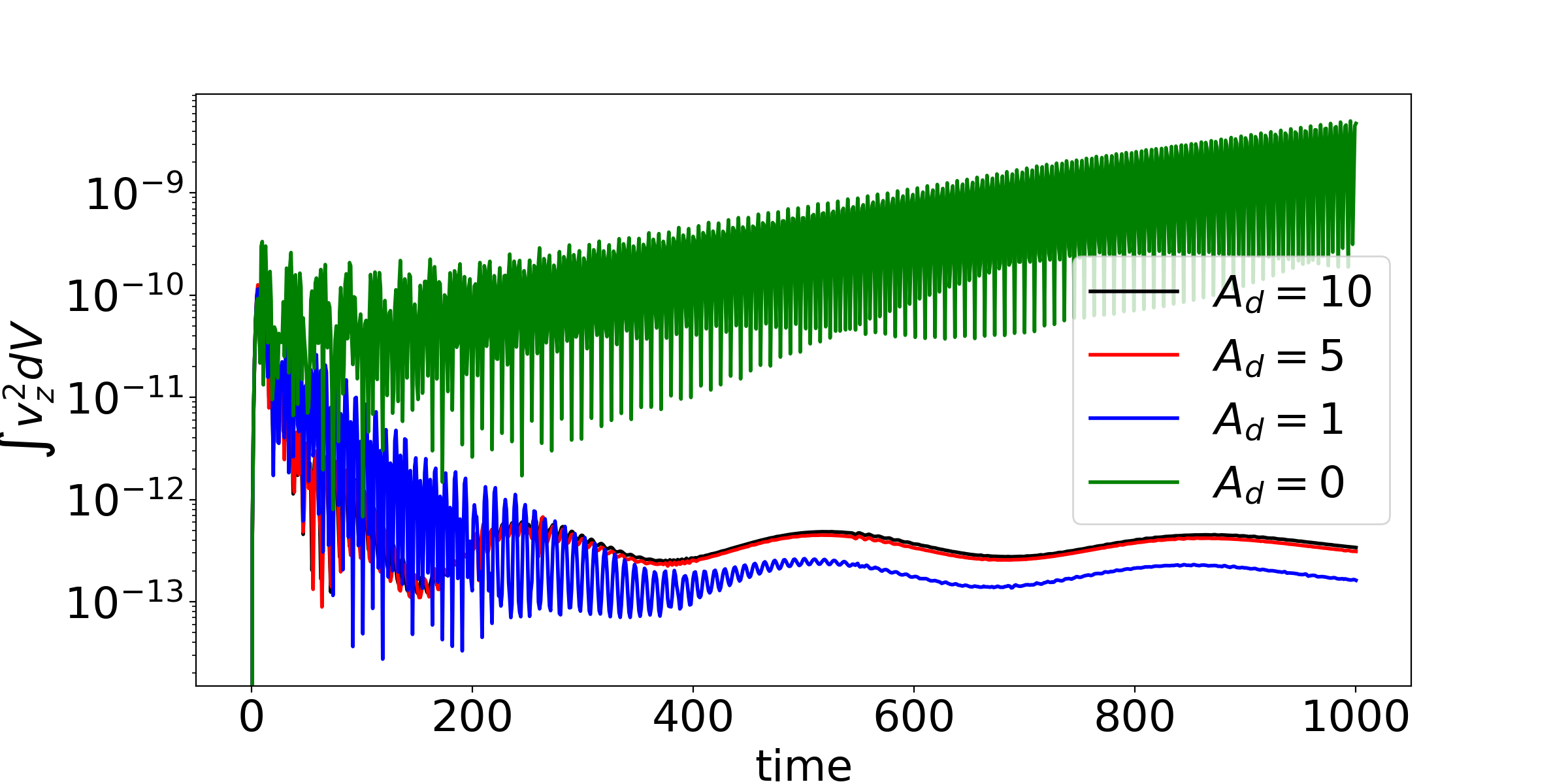}\\
\includegraphics[trim={0 1.7cm 3cm 1.65cm},clip,width=\linewidth]{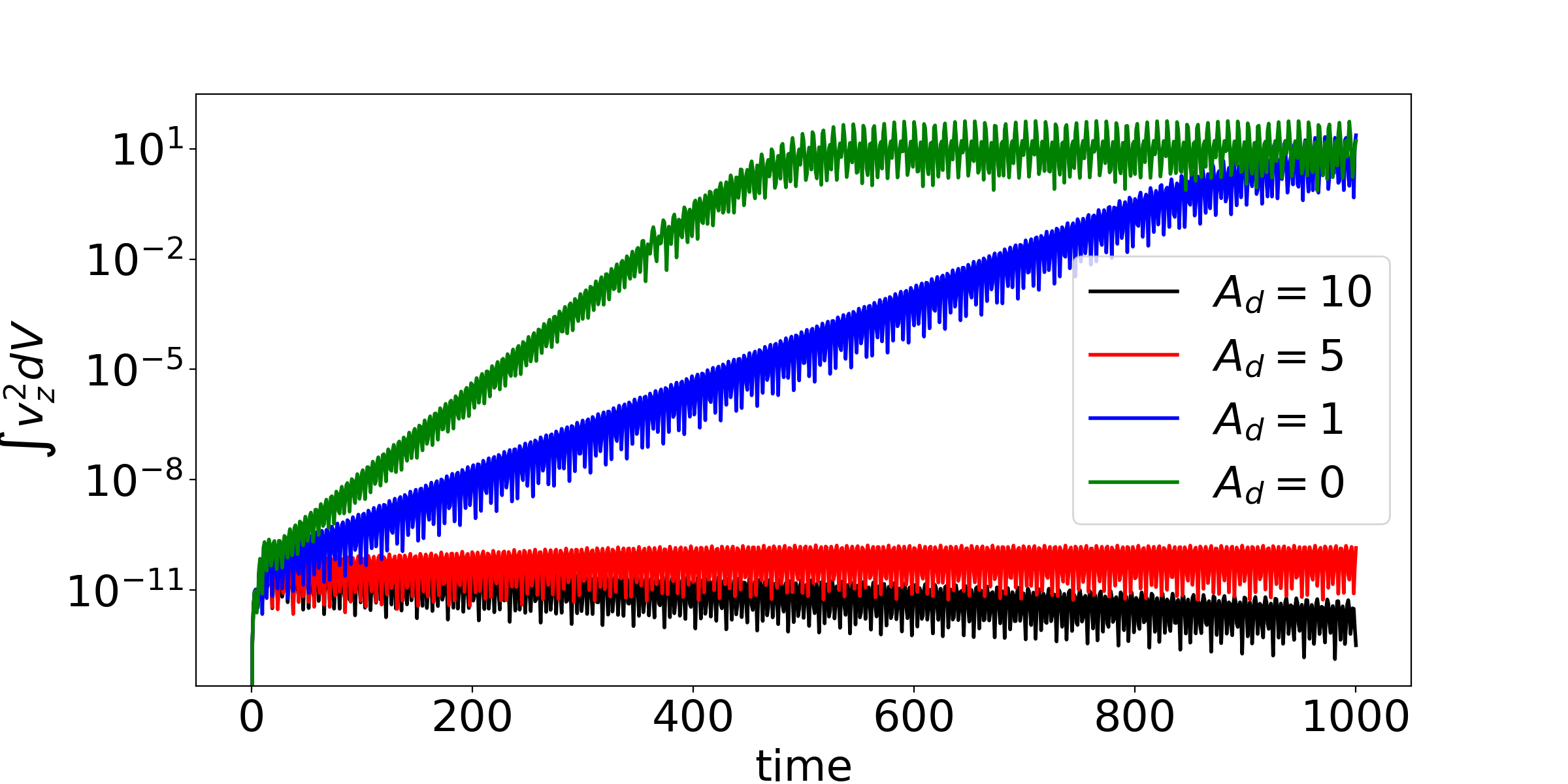}\\
\includegraphics[trim={0 0 3cm 1.65cm},clip,width=\linewidth]{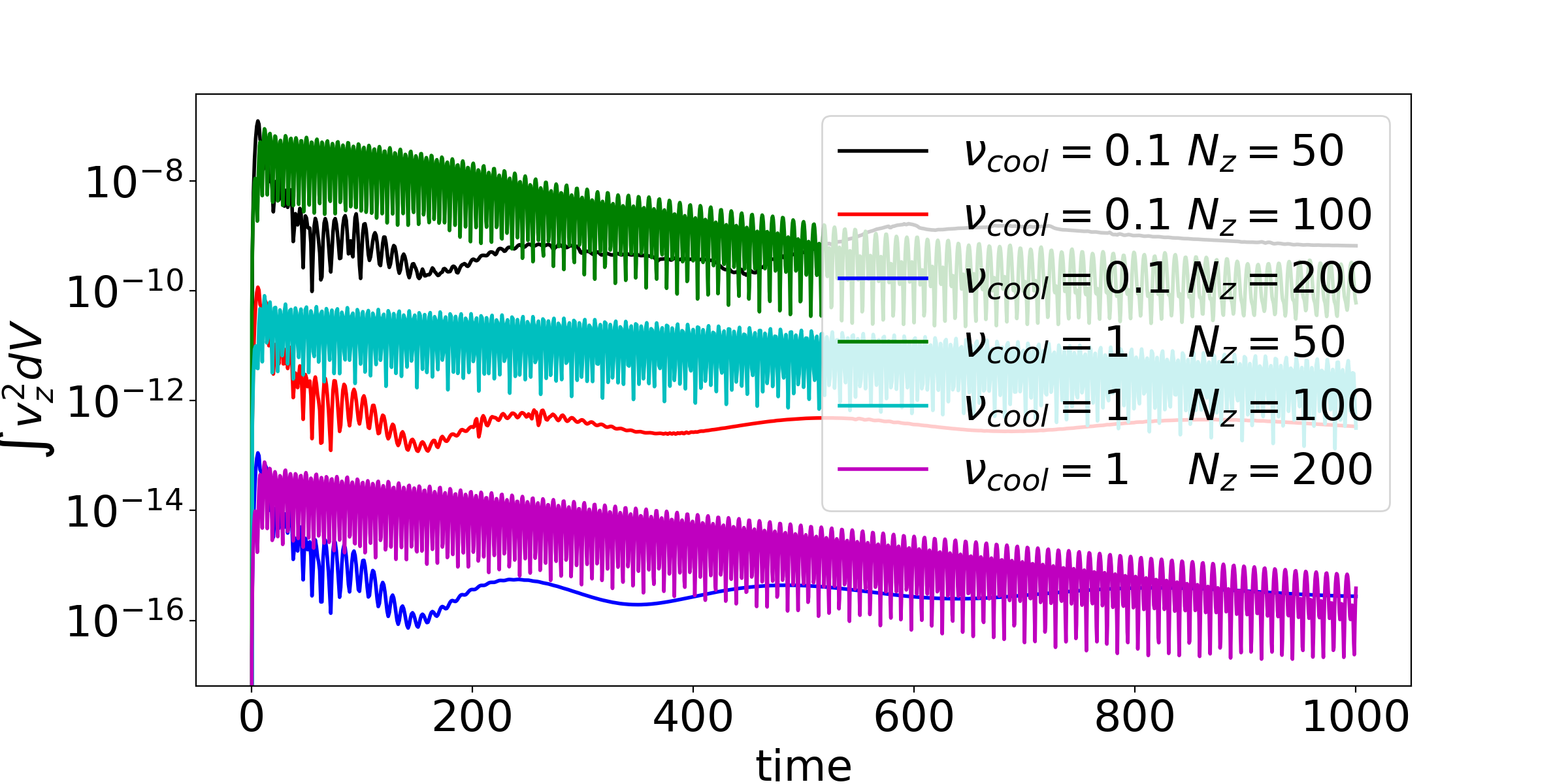}\\
\caption{Hydrostatic stability tests for variations of different parameters over the standard set $\nu_{\rm cool}=0.1$, $A_d=10$, $L=10$, $z_d=8$, $N_z=100$. From top to bottom, such variations are over (see legends for their values): $\nu_{\rm cool}$; combinations of domain size $\hat{L}$ and damping region position $z_d$ and symmetric radial boundary conditions on $v_z$ (purple) for $\nu_{\rm cool}=1$; $A_d$ with $\nu_{\rm cool}=0.1$; $A_d$ with $\nu_{\rm cool}=1$; resolution for two different $\nu_{\rm cool}$. We show the integrated L2 norm of $v_z$, which is the proxy to the discretisation error in case of no forcing (wind or perturbations) and no initial magnetic fields. The integral over the volume $V=(5 \times 5 \times 10)~H_*^3$ in units of $c_*^2 H_*^3$; time in units $t_*$.}
\label{fig:static_tests}
\end{figure}

In Fig.~\ref{fig:static_tests} we show the evolution of $\int \hat v_z^2 dV$ over $1000~t_*$, showing the variation of the different parameters. In the top panel, we see that any choice $\nu_{\rm cool} \lesssim 1$ provides similar and acceptable results, with velocities much lower than the $\nu_{\rm cool}=0$ case (which is low, but experiences a slow rise in $T$). Larger values of $\nu_{\rm cool} \gtrsim 2$ provide instead a persistent growth, that is unrelated to the timestep adopted.
In all cases, some oscillations are seen, reflecting some waves that are triggered by the initial readjustment. Such waves are damped more efficiently for intermediate values of $\nu_{\rm cool}$. The value $\nu_{\rm cool}=0.1$ adopted in the main text allows a damping in a few hundreds $t_*$.

In the second, third and fourth panels, we can see that cases with no damping ($L=8$, $A_d=0$) provide a continuous rise of the perturbation, which makes indeed the simulation unstable if we choose non-optimal values of other parameters (in particular, $\nu_{\rm cool}=1$, fourth panel). The oscillations are damped effectively and quickly for $A_d \gtrsim 5$. Using larger domains (cyan and green line in second panel), or shallower damping regions (blue) also produce the effect of keeping the oscillations longer and usually stronger (which again is more evident if we use other non-optimal values of $\nu_{\rm cool}=1$, as in the third panel). In the second panel we also show in purple the unstable case of other boundary conditions for $\hat v_z$ (i.e., $\hat v_z$ being flat close to the radial boundaries, instead of being zero): it makes any simulation highly unstable after few hundreds $t_*$ typically, causing inflows and outflows of matter regardless of the damping region.

These trends are valid for the other resolutions (see bottom panel), among which we see the expected scaling of the discretization error. These tests highlight the need of a damping region and to limit the size of the domain, due to the exponential background profile. After comparing tens of runs, we consider as optimal and standard the following set of parameters: $\nu_{\rm cool}=0.1$, $L=10$, $z_d=8$, $A_d=10$. Due to the damping region extension and the need to decrease the wind forcing $w(z)$ to zero at the top (see \S\ref{sec:forcing}), the useful physical region is limited to half of the domain, $\hat{z} \lesssim 5$.

\section{1D runs for winding only: initial magnetic field dependence}\label{app:initial_b}

\begin{figure}
\centering
\includegraphics[width=.9\linewidth]{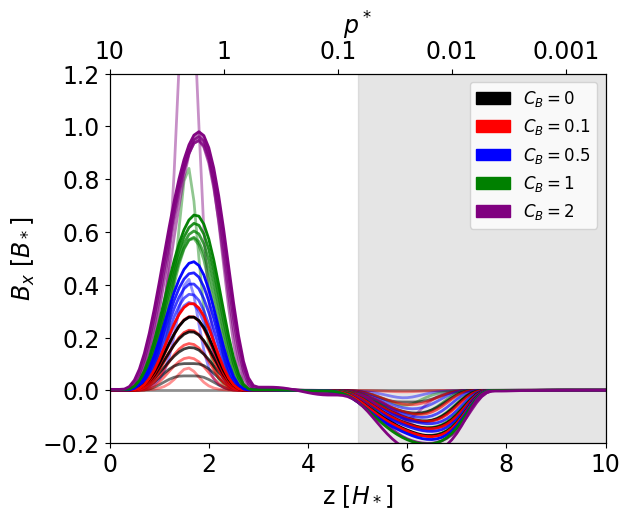}\\
\includegraphics[width=.9\linewidth]{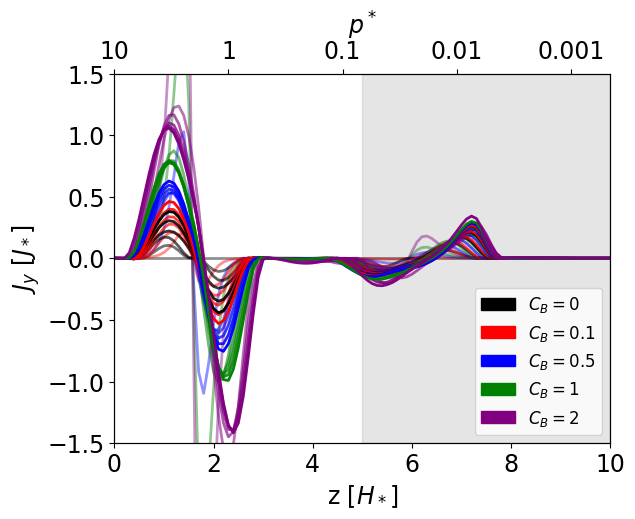}
\caption{Comparison of the vertical profile evolution of $B_x$ (top) and $J_y$ (bottom) in the 1D problem, i.e., in the absence of perturbations, $\lambda=0$. The four colors indicate different values of the initial magnetic field amplitude $C_B=\{0,0.1,0.5,1,2\}$, in the eq.~(\ref{eq:initial_b}). The five shades (transparent to opaque) indicate different times: $t/t_*=0,500,1000,1500,2000,2500$.}
\label{fig:initial_b_comp}
\end{figure}

In Fig.~\ref{fig:initial_b_comp} we show the comparison of the $B_x$ evolution between different 1D runs, i.e., without any perturbations, $\lambda=0$. Different colors mark different initial amplitudes of $B_x$ (eq.~\ref{eq:initial_b}), while the shades mark the evolution over long times ($t/t_*=0,500,1000,1500,2000,2500$). The initial profile (the most transparent lines) rapidly changes to a shape that resembles the asymptotic ones (the most opaque lines, at 2500 $t_*$), which is indeed similar to the full 3D cases discussed in the text, e.g. Fig.~\ref{fig:BxByBz}. Starting with low values of magnetic fields (black or red, $C_B=0,0.1$), the winding makes the field to grow, but it will take several thousands $t_*$ to approach the asymptotic profile. On the contrary, starting with a large value of $C_B$ (purple, $C_B=2$), the numerical dissipation damps the peaks and on average the field decreases. Moreover, in the latter case such value seems to be higher than in the other cases. The cases with $C_B=0.5,1$ (blue, green) reach instead much faster a similar asymptotic profile, which is also similar to the one approached by $C_B=0.1$. We also verified the same trends in the 3D: too small or too large values of $C_B$ hampers the ability of reach a quasi-asymptotic profile of the winding-generated $B_x$ profile. Therefore, we fix $C_B=1$ in our 3D simulations, keeping in mind that it can take up to $\sim 1000 ~ t_*$ to approach the asymptotic profile.

Note that such asymptotic profile depends on the wind profile and on the numerical diffusivity (see below).

\section{Numerical vs. physical diffusivity}\label{app:diffusivity}

\begin{figure}
\centering \includegraphics[width=\linewidth]{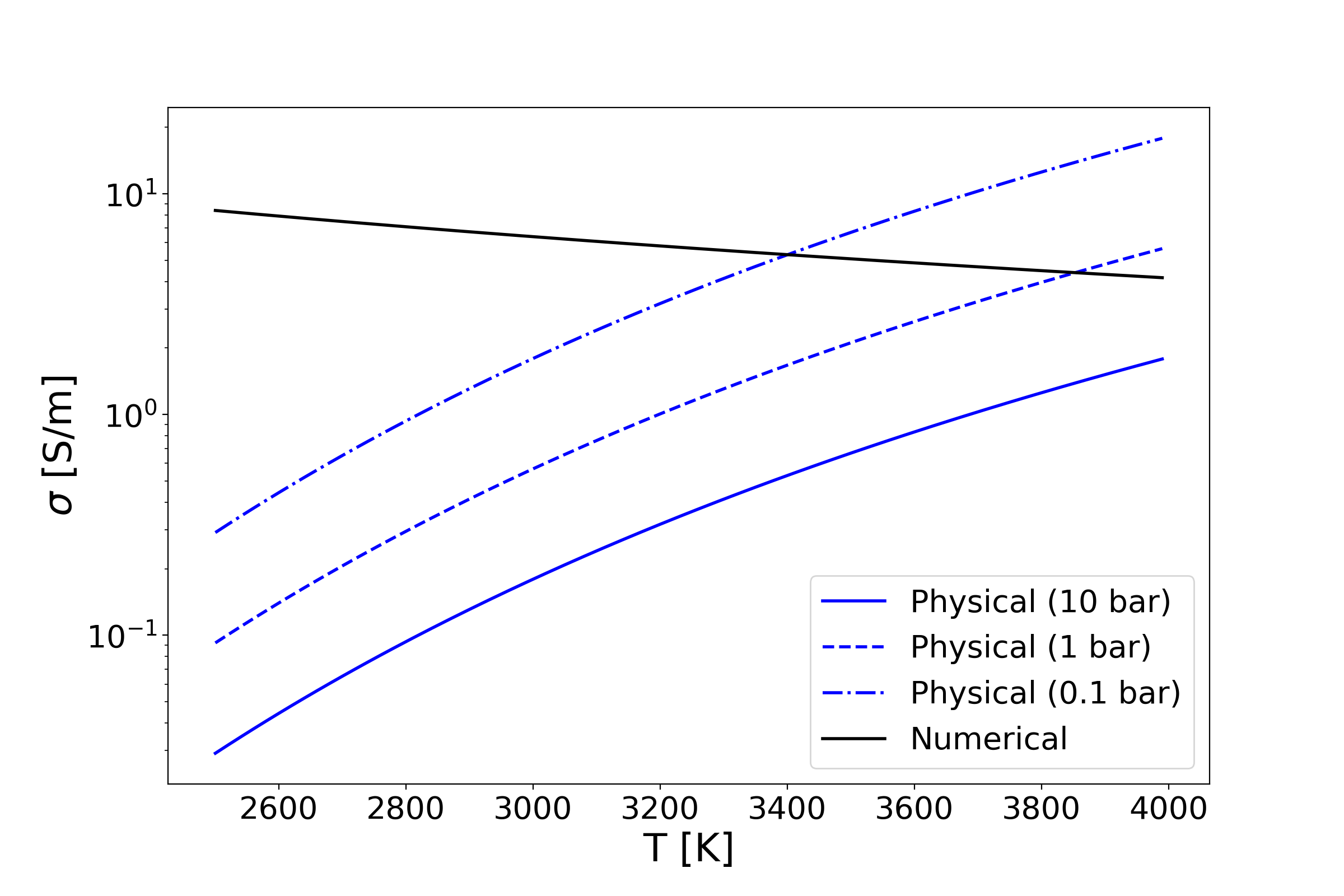}
\caption{Comparison between the physical conductivity $\sigma(T,p)$ as in eq.~(\ref{eq:sigma})  (blue lines, for $p=0.1,1,10$ bar) and numerical conductivity $\sigma_{\rm num}$ (estimated for $N_z=100$, black), for $g_{10}=\mu=1$, as a function of temperature. The results of this study are applicable if $\sigma_{\rm num}$ is not much larger than $\sigma$.}
\label{fig:conductivity}
\end{figure}

In Fig.~\ref{fig:conductivity} we compare the physical conductivity $\sigma(T,p)$, given by eq. (\ref{eq:sigma}), with the numerical conductivity $\sigma_{\rm num}(T):=1/(\eta_{\rm num}\mu_0)$. From this figure, we can determine the applicable region where $\sigma_{\rm num}$ is not much larger than $\sigma$. This is important especially in the region of the shear, $p\sim 1$ bar. Therefore, we conclude that: (i) the simulations presented here are valid for temperatures approaching $T\gtrsim 3800 K$ (for which $\sigma$ and $\sigma_{\rm num}$ are comparable in the shear layer region), and should give results not too far from the ones expected if resistivity is included, if $T\gtrsim 3000$ K (for which the difference between $\sigma$ at 1 bar and $\sigma_{\rm num}$ is less than one order of magnitude); (ii) higher resolutions would give numerical diffusivities that are too low, making the results unreliable.

\end{document}